\documentclass[preprint,preprintnumbers,amsmath,amssymb]{revtex4}
\usepackage[colorlinks=true, pdfstartview=FitV, linkcolor=blue, citecolor=red, urlcolor=magenta]{hyperref}
\usepackage{amssymb,amsmath,epsfig}
\allowdisplaybreaks

\begin{document}

\title{\bf Consequences of Thermal Fluctuations of Well-Known Black Holes in Modified Gravity}
\author{Abdul Jawad \footnote{jawadab181@yahoo.com;~~abduljawad@cuilahore.edu.pk}}
\address {Department of Mathematics, COMSATS University\\ Islamabad,
Lahore-Campus, Lahore-54000, Pakistan.}

\begin{abstract}
Quantum fluctuation consequences have significant role in
high-energy physics. These fluctuation often regarded as a
correction of the infrared (IR) limit. Such correction contribute to
the high-energy limit of thermodynamical quantities and the
stability conditions of black holes. In this work, we analyze the
thermal stability of black holes in the presence of thermal
fluctuations. We consider AdS black hole in Born-Infeld massive
gravity with non-abelian hair and the charged AdS black hole with a
global monopole. We develop many thermodynamical quantities such as
entropy, temperature, pressure, heat capacity of a system at
constant volume and pressure, ratio between the heat capacities at
constant pressure and volume, Gibbs free energy and Helmholtz free
energy for both black holes. The critical behavior and phase
transitions of black holes are also presented. We also observe the
local and global stability of black holes in the grand canonical
ensemble and canonical ensemble for the specific values of different
parameters, such as, symmetry breaking parameter $\eta$, massive
parameter $m$ and non-abelian hair $\nu$.
\end{abstract}

\maketitle

%\textbf{Keywords:}Thermodynamics. Regular black holes. Phase Transition. Thermal Stability.\\
%\textbf{PACS:} 95.36.+d; 98.80.-k.

\section{Introduction}

In GR, black holes (BHs) have shown many generative features of
spacetime such as the event horizon, the singularity, Hawking
radiation and their connection with thermodynamics. Among all of the
above phenomenon, the strong support to the thermodynamical
properties of BHs is the Hawking radiation. Classically, nothing can
escape from the event horizon of a BH while quantum mechanically it
has been proved that BH can emit radiation, known as Hawking
radiation \cite{2}-\cite{3} and has the Planck spectrum. Discovery
of Hawking radiation proved that BHs have temperature, therefore,
the concept which is proposed by Bekenstein about the entropy of BHs
is no longer a mystery. Also, the incredible work of Hawking made
the entropy quantitative which is related to the area of the event
horizon $S=\frac{A}{4}$ \cite{4}-\cite{5}. The BHs are fluctuating,
that is, the event horizon becomes fuzzy because of the quantized
metric. Because of the quantum fluctuations, the need of making
corrections to the maximum entropy of the BHs has been emerged,
which leads towards the development of the holographic principle,
which is an emerging new paradigm in quantum gravity
\cite{6}-\cite{7}. Moreover, the holographic principle implies that
the degrees of freedom in a spatial region can be encoded on its
boundary, with a density not exceeding one degree of freedom per
Planck cell \cite{8}.

Several approaches have been made practical and effective for
evaluating the entropy corrections such as by using the
non-perturbation quantum general relativity. In this approach the
density of microstates for asymptotically flat BHs have been
calculated which construct the logarithmic correction to the
standard Bekenstein entropy area relation \cite{9}. One can also use
the methodologies based on generating logarithmic correction terms
for all those BHs whose microscopic degrees of freedom are explained
by conformal field theory \cite{10}-\cite{11}, Hamiltonian partition
functions \cite{12}, loop quantum gravity \cite{13}, near horizon
symmetries \cite{14} and thermodynamic arguments \cite{15}.
Moreover, the corrections to the entropy of dilation BHs are
evaluated which emerged to be the logarithmic corrections \cite{16}.
By using the brick wall method, quantum corrections to the entropy
of a static spherically symmetric BH global monopole system arising
from the Dirac spinor field are investigated in Ref. \cite{17}. The
effect of thermal fluctuations for the entropy of both neutral and
charged black holes has been investigated in Ref. \cite{18}. Entropy
corrections for Schwarzschild BHs has been found by putting it in
the center of a spherical cavity of finite radius to achieve
equilibrium with surroundings \cite{19}. In a massive theory of
gravity, the effects of the quantum fluctuations has been analyzed
for a charged BTZ BH in asymptotically AdS and dS spacetimes
\cite{20}-\cite{21}. The effects of thermal fluctuations on charged
ADS BHs and modified Hayward BHs have been investigated in the
recent work \cite{22}-\cite{23}.

The dRGT model, which is one of the interesting theories of massive
gravity is introduced by the de Rham et al. \cite{24}-\cite{25}. On
the basis of the reference metric, there are various modifications
to the dRGT model and Vegh provide one of the good methods \cite{26}
and his model undergoes the breaking of the translational symmetry.
Zhang and Li proved that this model is stable \cite{27}. After that,
the thermodynamical properties and phase transition of many BHs have
been investigated \cite{28}-\cite{32}. Non-linear electrodynamics
(NED) theories are also very useful theories \cite{33}-\cite{41}. In
order to remove the divergency of self energy of a point-like
charge, Born and Infeld introduced one of the best and interesting
NED theory which is known as Born-Infeld (BI) theory \cite{42}. In
the past few years BI action is being used with the development of
superstring theory. D-branes dynamics and some soliton solutions of
super-gravity are governed by BI action. For many reasons, extension
of RN BH solutions in Einstein-Maxwell theory to charged BH
solutions in BI theory along a cosmological constant has attracted
some interest in past few years \cite{57}-\cite{58}. There are many
methods and approaches for studying and exploring the thermal
stability and phase transitions of BHs explained in \cite{59}.

Coupled to BI NED, BH solutions and their Van der Waals kind of
behavior in massive gravity has been investigated in \cite{44}. In
the Maxwell field, one assumes the non-abelian Yang-Mills (YM) field
coupled to gravity as a matter source. In Gauss-Bonnet-massive
gravity, the thermodynamical properties of BHs and their phase
transition in the existence of YM field have been investigated in
\cite{45}. In the presence of the YM and BI NED fields, the exact BH
solutions of Einstein-Massive theory has been obtained by Hendi and
Momennia \cite{46}. After getting motivation from the recent work of
Hendi and Momennia \cite{46a}, in which they study the thermodynamic
description of (a)dS black holes in Born-Infeld massive gravity with
a non-abelian hair. We extended their work to discuss the impact of
thermal corrections on different parameters of black holes. This
paper is outlined as follows: In Sec. \textbf{II}, we concentrate on
working out the thermodynamical quantities of the AdS BH in BI
massive gravity with a non-abelian hair in the existence of
logarithmic correction to entropy. Furthermore, in Sec.
\textbf{III}, for the charged AdS BH with a global monopole along
with the logarithmic correction to the entropy, we find the
conserved and thermodynamical quantities. Sec. \textbf{IV} contains
tables of the results and Sec. \textbf{V} is devoted for
conclusions.

\section{AdS Black Hole in BI Massive Gravity with a Non-Abelain Hair}

We consider following (3+1)-dimensional action of EYM-massive
gravity along with BI NED for the model
\begin{equation}\nonumber
{\L}_{G}=-\frac{1}{16\pi}\int_{M}d^{3+1}x\sqrt{-g}\bigg(R-2\Lambda+
\ell_{BI}(\digamma_{M})-\digamma_{YM}+m^{2}\sum_{i}c_{i}\emph{U}_{i}(g,f)\bigg),
\end{equation}
where $\ell_{BI}(\digamma_{M})$ and
$\digamma_{YM}=\texttt{Tr}(\digamma_{\mu\nu}^{(a)}\digamma^{(a)\mu\nu})$
are the Lagrangian of BI NED and YM invariant, $m$ is related to
graviton mass, $f$ make reference to an auxiliary metric, $c_{i}$
and $\emph{U}_{i}$ are free constants and symmetric polynomials of
$4\times4$ matrix
$\emph{K}_{\nu}^{\mu}=\sqrt{g^{\mu\sigma}f_{\sigma\nu}}$. We obtain
the three tensorial field equations which come from the variation of
action of the above equation w.r.t metric tensor $g_{\mu\nu}$,
Faraday tensor $F_{\mu\nu}$ and YM tensor $F_{\mu\nu}^{(a)}$, yield
as
\begin{equation}\nonumber
G_{\mu\nu}+\Lambda
g_{\mu\nu}=\texttt{T}_{\mu\nu}^{M}+\texttt{T}_{\mu\nu}^{YM}-m^{2}\chi_{\mu\nu},
\quad
\partial_{\mu}[\sqrt{-g}F^{\mu\nu}\partial_{\digamma}\ell_{BI}(\digamma)]=0,\quad
\hat{D}_{\mu}F^{(a)\mu\nu}=0,
\end{equation}
where $\hat{D}_{\mu}$ is covariant derivative of gauge field. The
energy-momentum tensor of electromagnetic and YM fields and
$\chi_{\mu\nu}$ is written as
\begin{eqnarray}\nonumber
\texttt{T}_{\mu\nu}^{M}&=&\frac{1}{2}g_{\mu\nu}\ell_{BI}(\digamma)-2\digamma_{\mu\nu}\digamma_{\nu}^{\lambda}\partial_{\digamma}\ell_{BI}(\digamma),
\\\nonumber
\texttt{T}_{\mu\nu}^{M}&=&\frac{1}{2}g_{\mu\nu}F_{\rho\sigma}^{(a)}F^{(a)\rho\sigma}+2F_{\mu\lambda}^{(a)}F_{\nu}^{(a)\lambda},
\\\nonumber
\chi_{\mu\nu}&=&-\frac{c_{1}}{2}(\emph{U}_{1}g_{\mu\nu}-\emph{K}_{\mu\nu})-\frac{c_{2}}{2}
(\emph{U}_{2}g_{\mu\nu}-2\emph{U}_{1}\emph{K}_{\mu\nu}+2\emph{K}_{\mu\nu}^{2})\\\nonumber&-&
\frac{c_{3}}{2}(\emph{U}_{3}g_{\mu\nu}-3\emph{U}_{2}\emph{K}_{\mu\nu}+6\emph{U}_{1}\emph{K}_{\mu\nu}^{2}-6\emph{K}_{\mu\nu}^{3})-\cdot\cdot\cdot
\end{eqnarray}

The exact AdS BH solution of Einstein-Massive theory in the presence
of YM and BI NED is being developed by Hendi and Momennia \cite{46}
and its line element is given by
\begin{equation}\label{1}
ds^{2}=-f(r)
dt^{2}+\frac{1}{f(r)}dr^{2}+r^{2}(d\theta^{2}+\sin^{2}d\phi^{2}),
\end{equation}
where the metric function is
\begin{equation}\label{2}
f(r)=1-\frac{m_{o}}{r}-\frac{\Lambda
r^{2}}{3}+\frac{\upsilon^{2}}{r^{2}}+\frac{m^{2}}{2}(cc_{1}r+2c^{2}c_{2})+\frac{2\beta^{2}r^{2}}{3}(1-\chi_{1}),
\end{equation}
and \begin{equation}\label{3}
\chi_{1}=_{2}\digamma_{1}(\frac{-1}{2},\frac{-3}{4};\frac{1}{4};\frac{-q^{2}}{\beta^{2}r^{4}}),
\end{equation}
which is a hypergeometric function, $\Lambda$ is the cosmological
constant, $q$ is the integration constant which is related to the
total electric charge of BH, $\nu$ is the magnetic parameter, $c,
~c_{1}$ and $c_{2}$ are free constants. Also $m_{o}$ is the
integration constant which relates with the total mass of BH. The
obtained solution possess Coulomb charge, massive term and a
non-abelian hair. In the obtained $f(r)$, fourth term
$\frac{\nu^{2}}{r^{2}}$ is related to the magnetic charge with
non-abelian hair, fifth term is related to the massive gravitons and
the last term is related to the non-linearity of electric charge.
For massless graviton $m=0$ and linear electrodynamics
$\beta\longrightarrow\infty$, $f(r)$ reduces to EYM solution with
Maxwell field. By setting $f(r=r_{+})=0$, we have
\begin{equation}\label{4}
m_{o}=r_{+}-\frac{\Lambda
r_{+}^{3}}{3}+\frac{\nu^{2}}{r_{+}}+\frac{m^{2}(cc_{1}r_{_{+}}^{2}+2c^{2}c_{2}r_{+})}{2}+\frac{2\beta^{2}r_{+}^{3}(1-\chi_{1})}{3}.
\end{equation}
Using the Hamiltonian approach, it was introduced that the total
mass $M$ of BH can be obtained by the massive gravity as \cite{47}
\begin{equation}\label{5}
M=\frac{m_{o}}{2}.
\end{equation}
Hence, the total mass of BH becomes
\begin{equation}\label{6}
M=\frac{r_{+}}{2}-\frac{\Lambda
r_{+}^{3}}{6}+\frac{\nu^{2}}{2r_{+}}+\frac{m^{2}(cc_{1}r_{+}^{2}+2c^{2}c_{2}r_{+})}{4}+\frac{\beta^{2}r_{+}^{3}(1-\chi_{1})}{3},
\end{equation}
which implies that outer horizon $r_{+}\neq0$. The entropy and
volume is related to BH horizon defined as \cite{23}
\begin{equation}\label{7}
S_{o}=\pi r_{+}^{2},~~V=\frac{4\pi r_{+}^{3}}{3}.
\end{equation}
%\textbf{The expression for volume $V$ in Eq. (\ref(7)) is same as
%the Eq. \textbf{10} in \cite{66}}.
In order to study the thermal fluctuations, thermal stability and
phase transitions of BHs, we examine the conserved and thermodynamic
quantities such as pressure, entropy, specific heats, Gibbs free
energy and Helmholtz free energy of BHs by using the logarithmic
correction terms. The temperature of AdS BH in BI massive gravity
with non-abelian hair can be written as
\begin{equation}\label{9}
T=\frac{f'(r)}{4\pi}\mid_{r=r_{+}}=\frac{1}{4\pi
r_{+}}\bigg(1-\Lambda
r^{2}_{+}-\frac{\nu^{2}}{r^{2}_{+}}+m^{2}(cc_{1}r_{+}+c^{2}c_{2})
+2\beta^{2}r^{2}_{+}\bigg(1-\sqrt{1+\frac{q^{2}}{\beta^{2}r^{2}_{+}}}\bigg)
\bigg).
\end{equation}
Now we use the logarithmic correction terms for the entropy in order
to discuss the thermal fluctuations. The corrected term for the
entropy turns out to be \cite{48}
\begin{equation}\label{11}
S=S_{o}-\frac{b}{2}\log[S_{o}T^{2}],
\end{equation}
where $b$ is a constant parameter, which is added in order to handle
the logarithmic correction terms produce because of the thermal
fluctuations. By setting $b=0$, entropy can be recovered without any
correction term.

Using Eqs. (\ref{7}) and (\ref{9}) in Eq. (\ref{11}), we obtain
\begin{equation}\label{12}
S=\pi r_{+}^{2}-\frac{b}{2}\log\bigg[\frac{\bigg(1-\Lambda
r_{+}^{2}-\frac{\nu^{2}}{r_{+}^{2}}+m^{2}(cc_{1}r_{+}+c^{2}c_{2})+2\beta^{2}
r_{+}^{2}\bigg(1-\sqrt{1+\frac{q^{2}}{\beta^{2}r_{+}^{2}}}\bigg)^{2}}{16\pi}\bigg].
\end{equation}
\begin{figure} \centering
\epsfig{file=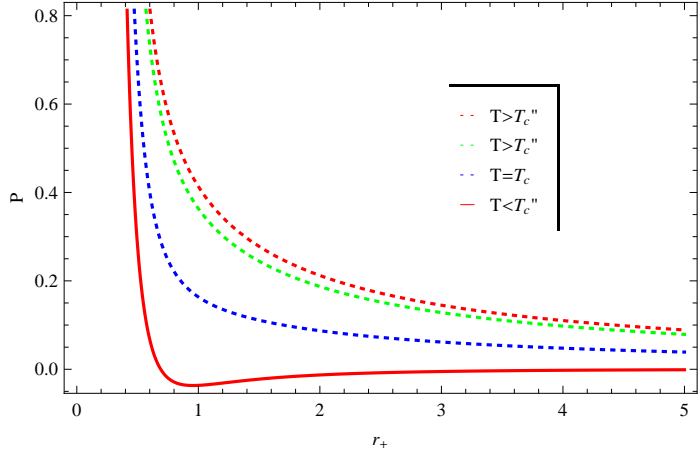,width=.60\linewidth}\caption{Plot of $P$ versus
$r_{+}$ for AdS BH in BI massive gravity with a non-abelain hair.
Specific values of different parameters are $\beta=1, ~\nu=1, ~c=1$,
$m=1$ and $c_{2}=2$.}
\end{figure}
We consider usual thermodynamical pressure of BHs defined as
\cite{66},
\begin{equation}\label{14}
P=\frac{-\Lambda}{8\pi}.
\end{equation}
Using Eqs. (\ref{9}) and (\ref{14}), one can obtain the equation of
state for the considered BH as follow
\begin{equation}\label{14a}
P=-\frac{r^2+2 \beta^2 r^4-2 \beta^2 \sqrt{1+\frac{q^2}{\beta^2
r^2}} r^4-4 \pi r^3 T-v^2+c m^2 r^3 c_1+c^2 m^2 r^2 c_2}{8 \pi
r^4}.
\end{equation}
%Thus, these quantities satisfy the first law of thermodynamics.
FIG.\textbf{1} demonstrates $P$ versus $r_{+}$ for AdS BH in BI
massive gravity with a non-abelain hair. The two upper dotted red
and green lines correspond to the ``ideal gas" phase for $T
> T_{c}$, the critical isotherm $T = T_{c}$ is denoted by the dotted
blue line, lower solid line corresponds to temperatures smaller than
the critical temperature.

\subsection{Thermal Stability}

In BH thermodynamics, the amount of heat required to change the
temperature of a BH, is known as thermal capacity or heat capacity.
There are two types of heat capacities, one which measures the
specific heat when heat is added to the system at the constant
pressure $C_{_{p}}$ and the other which measure the specific heat
when the heat is added to the system at the constant volume $C_{v}$.
We obtain the specific heat at the constant volume by using the
following relation
\begin{equation}\label{17}
C_{v}=T(\frac{\partial S}{\partial r})(\frac{\partial r}{\partial
T}).
\end{equation}
Using Eqs.(\ref{9}) and (\ref{12}), we obtain
\begin{eqnarray}\nonumber
C_{v}&=&\frac{-1}{r_{+}\bigg(-\frac{4q^{2}}{r_{+}^{4}\sqrt{1+\frac{q^{2}}{\beta^{2}r_{+}^{4}}}}
-2\beta^{2}+2\beta^{2}\sqrt{1+\frac{q^{2}}{\beta^{2}r_{+}^{4}}}+\Lambda-\frac{3\nu}{r_{+}^{4}}
+\frac{1+c^{2}m^{2}c_{2}}{r_{+}^{2}}\bigg)}\bigg(1-2r_{+}^{2}\bigg(\\\nonumber&\times&-1
+\sqrt{1+\frac{q^{2}}{\beta^{2}r_{+}^{4}}}\beta^{2}-r_{+}^{2}\Lambda-\frac{\nu}{r_{+}^{2}}
+cm^{2}(r_{+}c_{1}+cc_{2})\bigg)\bigg)\bigg(2\pi
r_{+}-\bigg(
b\bigg(4r_{+}\beta^{2}\\\nonumber&-&\frac{4r_{+}\beta^{2}}{\sqrt{1+\frac{q^{2}}{\beta^{2}r_{+}^{4}}}}
-2r_{+}\Lambda+\frac{2\nu}{r_{+}^{3}}+cm^{2}c_{1}\bigg)\bigg)\bigg(1-2r_{+}^{2}\bigg(-1+\sqrt{1
+\frac{q^{2}}{\beta^{2}r_{+}^{4}}}\beta^{2}-r_{+}^{2}\Lambda\\\label{18}&-&\frac{\nu}{r_{+}^{2}}
+cm^{2}(r_{+}c_{1}+cc_{2})\bigg)\bigg)^{-1}\bigg).
\end{eqnarray}
The behaviors of heat capacity versus event horizon of the AdS BH in
BI massive gravity with a non-abelian hair is displayed in FIGs.
\textbf{2} and \textbf{3}. In order to check the local stability of
BH, we discuss two cases for fixed values of $m$. When $\Lambda=-1$
and $m=1$, the heat capacity always remain positive and indicates
that BH is locally stable and no phase transitions take place. For
$m=2$, there occur divergence point at $r_{+}=2.9$ and for
$r_{+}<2.9$ BH becomes unstable. While for $r_{+}>2.9$, BH becomes
thermodynamically stable. For $m=3$, divergence point occurs at
$r_{+}=4.4$ and the phase transitions take place between small BH
(SBH) and large BH (LBH). However, BH becomes locally stable for
$r_{+}>5.75$, for $\Lambda=1$ and all fixed values of $m$, heat
capacity is positive implying that no phase transition will take
place and thus, the BH is locally stable for $\Lambda=1$.
FIGs.\textbf{4} and \textbf{5} represent the heat capacity versus
event horizon of BH for fixed values of $\nu$ for negative and
positive cosmological constant. For the case of negative
cosmological constant heat capacity is negative for all the fixed
values throughout the region which show the local usability of BH.
For positive cosmological constant, there are stable and unstable
regions depend upon the values of $\nu$, as shown in FIG.\textbf{5}.
\begin{figure}
\begin{minipage}{14pc}
\includegraphics[width=18pc]{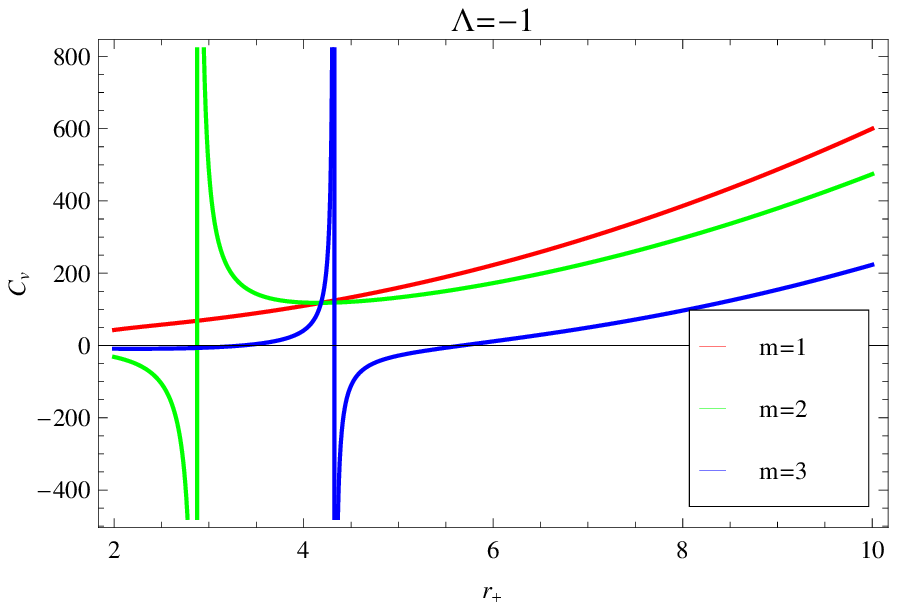}
\caption{\label{label}Plot of specific heat versus horizon radius
for negative cosmological constant. Specific values of parameters
are $\beta=1, ~\nu=1, ~c=1, c_{1}=1, c_{2}=1$ and $b=1$}
\end{minipage}\hspace{3pc}%
\begin{minipage}{14pc}
\includegraphics[width=18pc]{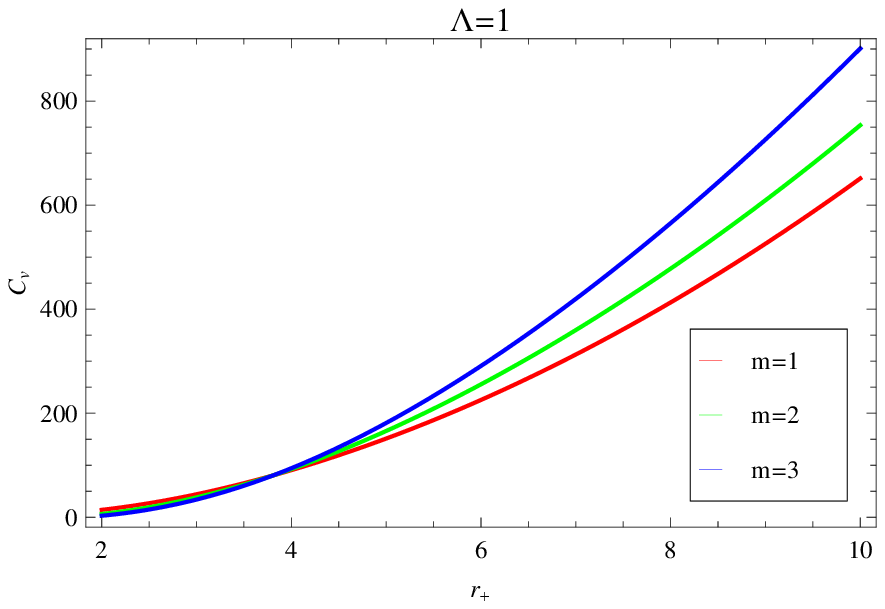}
\caption{\label{label}Plot of specific heat versus horizon radius
for positive cosmological constant. Specific values of parameters
are $\beta=1, ~\nu=1, ~c=1, c_{1}=1, c_{2}=1$ and $b=1$}
\end{minipage}\hspace{3pc}%
\end{figure}
\begin{figure}
\begin{minipage}{14pc}
\includegraphics[width=18pc]{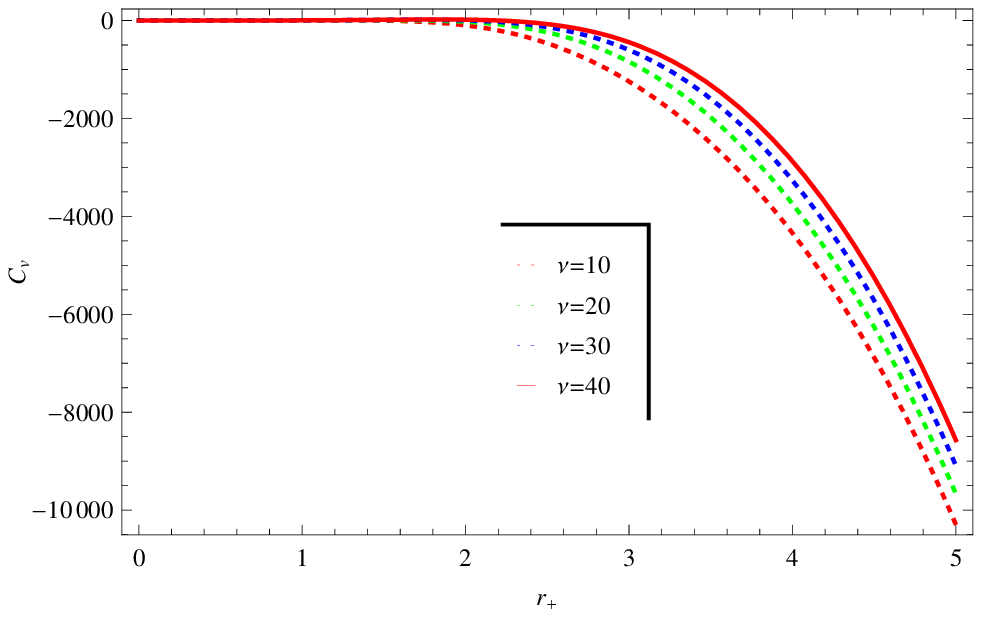}
\caption{\label{label}Plot of specific heat versus horizon radius
for negative cosmological constant. Specific values of parameters
are $\beta=1, ~c=1, c_{1}=1, c_{2}=1$ and $b=1$}
\end{minipage}\hspace{3pc}%
\begin{minipage}{14pc}
\includegraphics[width=18pc]{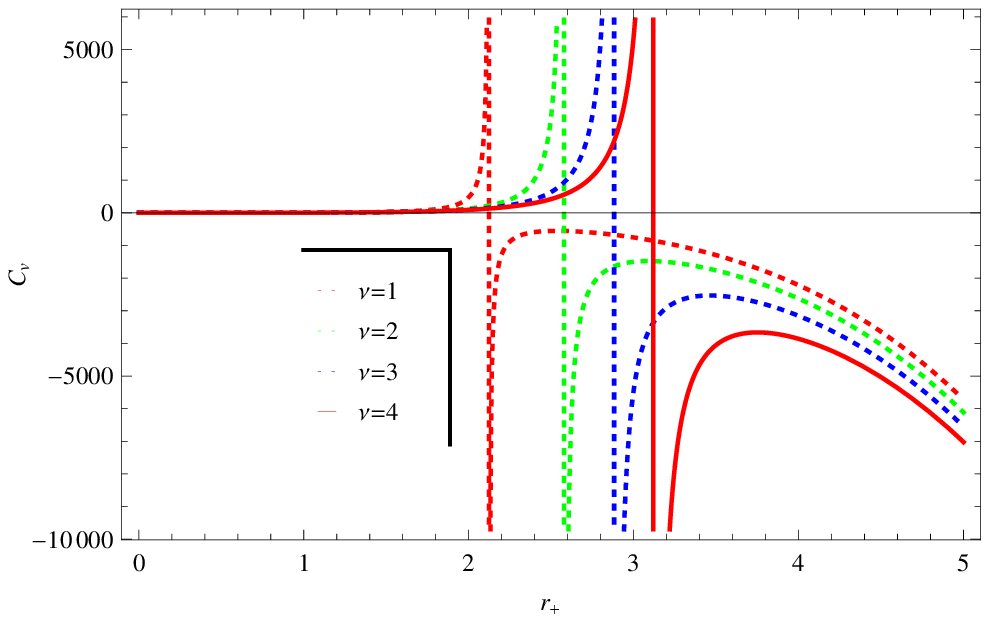}
\caption{\label{label}Plot of specific heat versus horizon radius
for positive cosmological constant. Specific values of parameters
are $\beta=1, ~c=1, c_{1}=1, c_{2}=1$ and $b=1$}
\end{minipage}\hspace{3pc}%
\end{figure}
Furthermore, the specific heat at the constant pressure can be
evaluated by the following relation
\begin{equation}\label{19}
C_{p}=\frac{(\frac{\partial M}{\partial r})}{(\frac{\partial
T}{\partial r})}.
\end{equation}
In view of Eqs. (\ref{6}) and (\ref{12}), we have
\begin{equation}\label{20}
C_{p}=\frac{-2\pi\bigg(r_{+}^{2}-r_{+}^{4}(2\beta^{2}(-1+\sqrt{1-\frac{q^{2}}{\beta^{2}
r_{+}^{4}}})+\Lambda)-\nu^{2}+cm^{2}r_{+}^{2}(r_{+}c_{1}+cc_{2})\bigg)}
{r_{+}^{2}\bigg(-\frac{4q^{2}}{r_{+}^{4}\sqrt{1+\frac{q^{2}}{\beta^{2}
r_{+}^{4}}}}-2\beta^{2}+2\beta^{2}\sqrt{1+\frac{q^{2}}{\beta^{2}r_{+}^{4}}}
+\Lambda-\frac{3\nu}{r_{+}^{4}}+\frac{1+c^{2}m^{2}c_{2}}{r_{+}^{2}}\bigg)}.
\end{equation}
The ratio of the above two heat capacities is also investigated
which can be expressed as
\begin{equation}\label{21}
\gamma=\frac{C_{p}}{C_{v}}.
\end{equation}
\begin{figure}
\begin{minipage}{14pc}
\includegraphics[width=18pc]{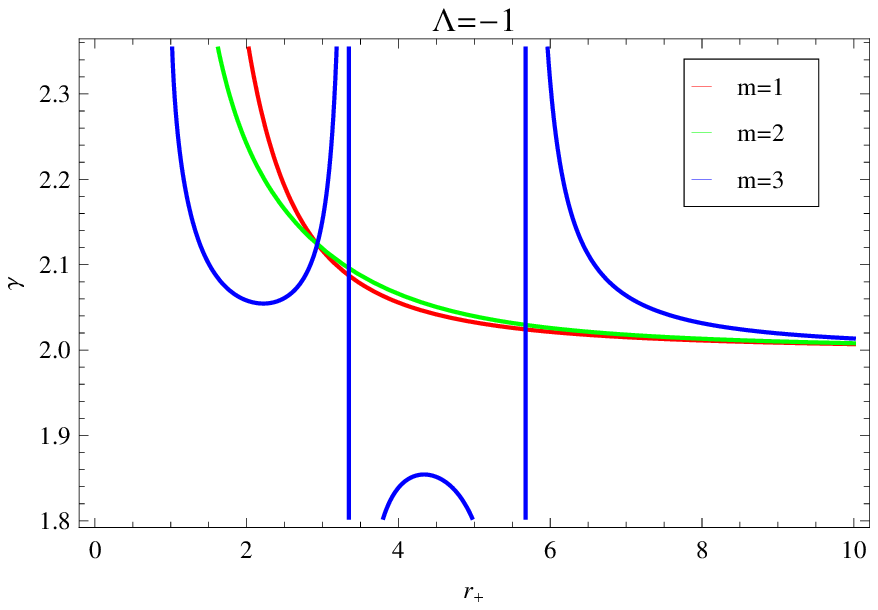}
\caption{\label{label}Plot of $\gamma$ versus $r_{+}$ for AdS BH in
BI massive gravity with a non-abelian hair for negative cosmological
constant. Specific values of different parameters are $\beta=1,
~\nu=1, ~c=1, ~c_{1}=-1, ~c_{2}=2$ and $b=1$}
\end{minipage}\hspace{3pc}%
\begin{minipage}{14pc}
\includegraphics[width=18pc]{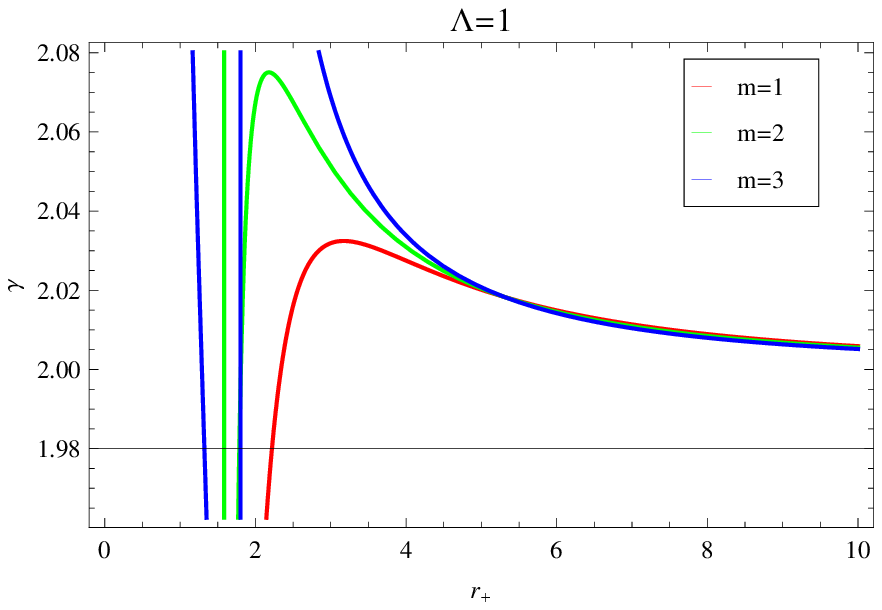}
\caption{\label{label}Plot of $\gamma$ versus $r_{+}$ for AdS BH in
BI massive gravity with a non-abelian hair for positive cosmological
constant. Specific values of parameters are $\beta=1, ~\nu=1, ~c=1,
~c_{1}=-1, ~c_{2}=2$ and $b=1$}
\end{minipage}\hspace{3pc}%
\end{figure}
Using Eqs. (\ref{18}) and (\ref{20}), we obtain
\begin{eqnarray}\label{22}
\gamma&=&\bigg(4\pi
r_{+}^{2}(q^{2}+r_{+}^{4}\beta^{2})\big(-r_{+}^{2}+r_{+}^{4}(2\beta^{2}(-1+\sqrt{1-\frac{q^{2}}{\beta^{2}r_{+}^{4}}})+\Lambda)+\nu^{2}-c
m^{2}r_{+}^{2}(r_{+}c_{1}\\\nonumber&+&cc_{2})\big)\bigg)
\bigg(-4b\beta^{4}r_{+}^{8}\sqrt{1+\frac{q^{2}}{\beta^{2}r_{+}^{4}}}+2b(q^{2}+r_{+}^{4}\beta^{2})(r^{4}(2\beta^{2}-\Lambda)+\nu)+2\pi
r_{+}^{2}(q^{2}+r_{+}^{4}\\\nonumber&+&\beta^{2})\big(-r_{+}^{2}+r_{+}^{4}\big(2\beta^{2}\big(-1+\sqrt{1+\frac{q^{2}}{\beta^{2}r_{+}^{4}}}\big)+\Lambda\big)+\nu\big)+cm^{2}r_{+}^{3}(q^{2}+r_{+}^{4}\beta^{2})\big((b-2\\\nonumber&\times&\pi
r_{+}^{2})c_{1}-2c\pi r_{+}c_{2}\big)\bigg)^{-1}.
\end{eqnarray}
FIGs. \textbf{6} and \textbf{7} illustrate the behavior of $\gamma$
for $\Lambda=-1$ and $\Lambda=1$ respectively. In the presence of
logarithmic correction to the entropy, the value of the $\gamma$
increases for $\Lambda=1$ and decreases for $\Lambda=-1$. When
$\Lambda=-1$, the value of $\gamma$ decreases for $m=1$ and $m=2$
while for $m=3$, the values of $\gamma$ has discontinuities. When
$\Lambda=1$, the values of $\gamma$ increases for small horizon and
then decrease to same values.

\subsection{The Gibbs free energy}

In order to analyze the global stability of BH, we now treat AdS BH
in BI massive gravity with a non-abelian hair as a thermodynamical
object by considering it in a grand canonical ensemble. In the grand
canonical ensemble free energy is also known as the Gibbs free
energy and it is defined as
\begin{equation}\label{23}
G=M-TS+PV.
\end{equation}
Inserting Eqs. (\ref{6}), (\ref{7}), (\ref{9}), (\ref{12}) and
(\ref{14a}) into (\ref{23}), we get
\begin{eqnarray}\nonumber
G&=&r_{+}-\frac{r_{+}^{3}\Lambda}{3}+\frac{\nu^{2}}{r_{+}}-\frac{2r_{+}^{3}\beta^{2}}{3}
\bigg(-1+_{2}\digamma_{1}\big(\frac{-1}{2},\frac{-3}{4};\frac{1}{4};\frac{-q^{2}}
{\beta^{2}r_{+}^{4}}\big)\bigg)+\frac{1}{2}cm^{2}r_{+}\big(r_{+}c_{1}+2cc_{2}\big)\\\nonumber&+&\bigg(\frac{1}{8\pi
r_{+}^{3}}\bigg)\times\bigg(2\pi
r_{+}^{2}+b\log[16\pi]-b\log[\bigg(-1+2r_{+}^{2}\beta^{2}\bigg(-1+\sqrt{1+\frac{q^{2}}
{r_{+}^{4}\beta^{2}}}\bigg)+r_{+}^{2}\Lambda\\\nonumber&+&\frac{\nu}{r_{+}^{2}}
-cm^{2}(r_{+}c_{1}+cc_{2})\bigg)^{2}]\bigg)\bigg(-r_{+}^{2}+r_{+}^{4}
\bigg(2\beta^{2}\bigg(-1+\sqrt{1+\frac{q^{2}}{r_{+}^{4}\beta^{2}}}\bigg)
+\Lambda\bigg)\\\nonumber&+&\nu-cm^{2}r_{+}^{2}(r_{+}c_{1}+cc_{2})\bigg)-\frac{r^{3}}{6
}((-12 M r+6 r^2+6 v^2+4 r^4 \beta ^2-4 r^4 \chi_{1} \beta ^2+3 c
\\\label{24}&\times&m^2 r^3 c_1+6 c^2 m^2 r^2 c_2)(2 r^4)^{-1})
\end{eqnarray}
\begin{figure}
\begin{minipage}{14pc}
\includegraphics[width=18pc]{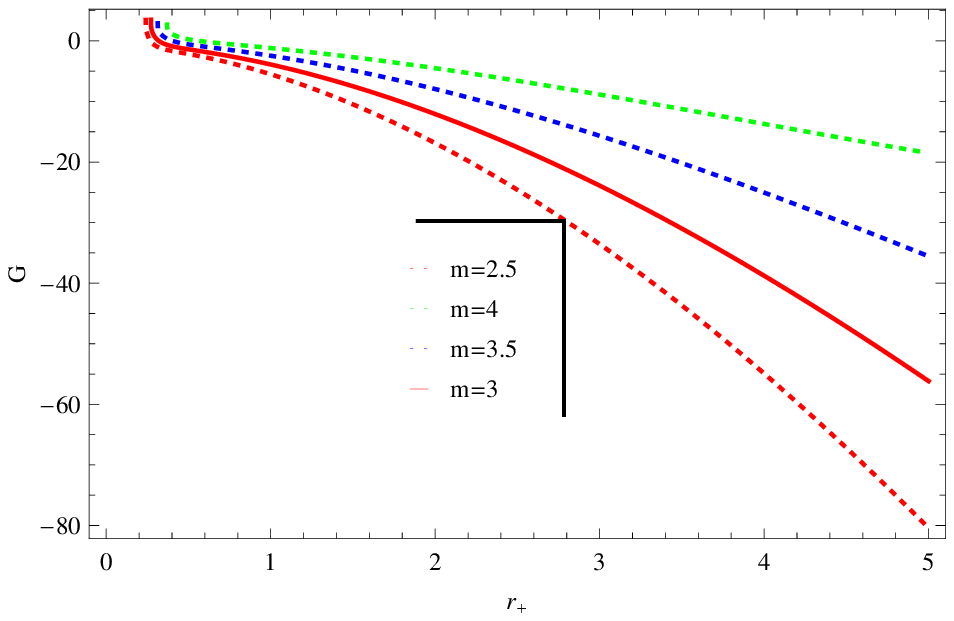}
\caption{\label{label}Plot of Gibbs free energy versus $r_{+}$ for
AdS BH in BI massive gravity with a non-abelian hair for negative
cosmological constant. Specific values of parameters are $\beta=1,
~\nu=1, ~c=1, ~c_{1}=1, ~c_{2}=2$ and $b=1$}
\end{minipage}\hspace{3pc}%
\begin{minipage}{14pc}
\includegraphics[width=18pc]{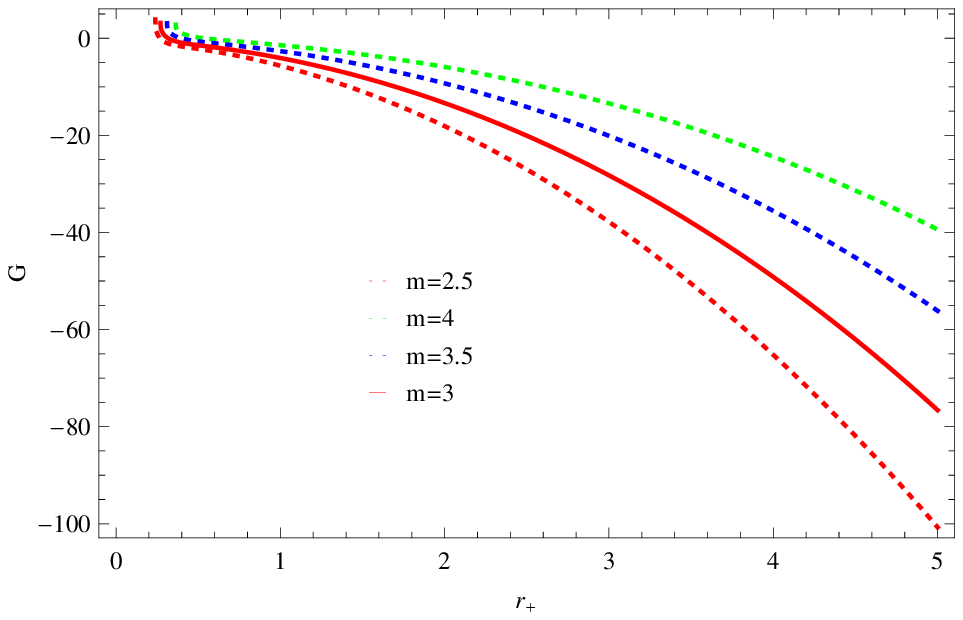}
\caption{\label{label}Plot of Gibbs free energy versus $r_{+}$ for
AdS BH in BI massive gravity with a non-abelian hair for positive
cosmological constant. Specific values of parameters are $\beta=1,
~\nu=1, ~c=1, ~c_{1}=1, ~c_{2}=2$ and $b=1$}
\end{minipage}\hspace{3pc}%
\end{figure}

The graphical representation of the Gibbs free energy and the
horizon radius for the fixed values of massless graviton and
cosmological constant takes place in the FIGs. \textbf{8} and
\textbf{9}. The left panel is for $\Lambda=-1$, while right panel is
for $\Lambda=1$, one can see that the Gibbs free energy is negative
throughout the region for both the cases. Gibbs free energy is used
to discuss the global stability of BH. The negative trajectories for
both the mentioned cases for all the values of $m$ represent the
unstable region.

\subsection{Canonical ensemble}

If the transference of charge on the BH is restrained then the BH
could be regarded as a thermodynamically closed system such as a
conical ensemble. When the charge is fixed, the free energy is known
as the Helmholtz free energy in the canonical ensemble and its
expression can be derived by using the following relation
\begin{equation}\label{25}
F=G-PV.
\end{equation}
By using Eq. (\ref{24}), the Helmholtz free energy for AdS BH in BI
massive gravity with a non-abelain hair can be obtained as
\begin{eqnarray}\nonumber
F&=&r_{+}-\frac{r_{+}^{3}\Lambda}{3}+\frac{\nu^{2}}{r_{+}}-\frac{2r_{+}^{3}
\beta^{2}}{3}\bigg(-1+_{2}\digamma_{1}\big(\frac{-1}{2},\frac{-3}{4};\frac{1}{4}
;\frac{-q^{2}}{\beta^{2}r_{+}^{4}}\big)\bigg)+\frac{1}{2}cm^{2}r_{+}
\big(r_{+}c_{1}+2cc_{2}\big)\\\nonumber&+&\bigg(\frac{1}{8\pi
r_{+}^{3}}\bigg)\bigg(2\pi
r_{+}^{2}+b\log[16\pi]-b\log\bigg[\bigg(-1+2r_{+}^{2}\beta^{2}\bigg(-1+\sqrt{1
+\frac{q^{2}}{r_{+}^{4}\beta^{2}}}\bigg)+r_{+}^{2}\Lambda\\\nonumber&+&\frac{\nu}
{r_{+}^{2}}-cm^{2}(r_{+}c_{1}+cc_{2})\bigg)^{2}\bigg]\bigg)\bigg(-r_{+}^{2}
+r_{+}^{4}\bigg(2\beta^{2}\bigg(-1+\sqrt{1+\frac{q^{2}}{r_{+}^{4}\beta^{2}}}\bigg)
+\Lambda\bigg)+\nu-cm^{2}\\\nonumber&\times&r_{+}^{2}(r_{+}c_{1}+cc_{2})\bigg)
+(16\pi^{2}r_{+}^{3})\bigg(3\bigg(-\frac{4q^{2}}{r_{+}^{4}\sqrt{1+\frac{q^{2}}
{\beta^{2}r_{+}^{4}}}}-2\beta^{2}+2\beta^{2}\sqrt{1+\frac{q^{2}}{\beta^{2}r_{+}^{4}}}
+\Lambda-\frac{3\nu}{r_{+}^{4}}\\\label{26}&+&\frac{1+c^{2}m^{2}c_{2}}{r_{+}^{2}}\bigg)\bigg)^{-1}.
\end{eqnarray}
\begin{figure}
\begin{minipage}{14pc}
\includegraphics[width=18pc]{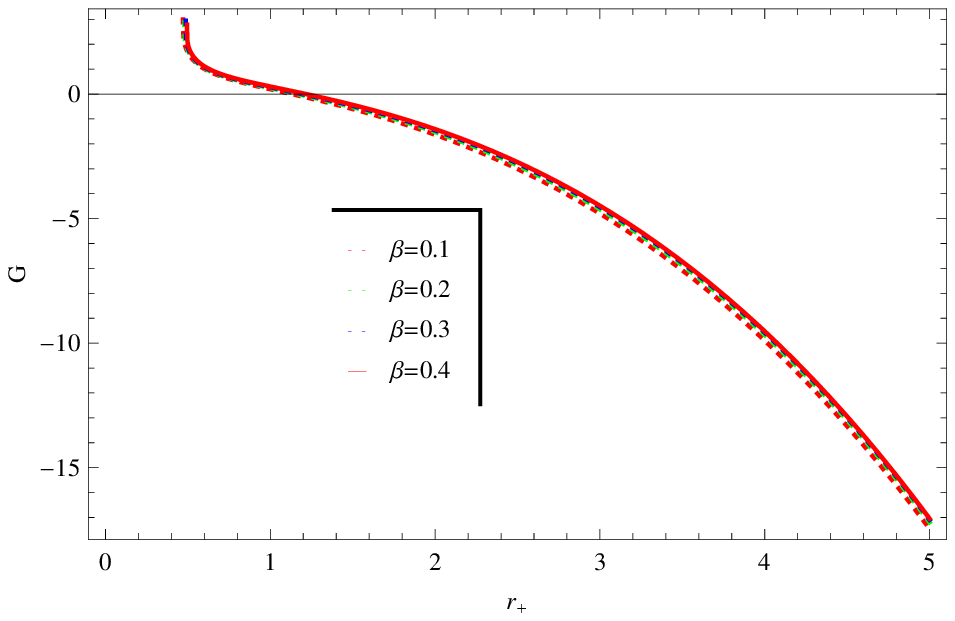}
\caption{\label{label}Plot of Gibbs free energy versus $r_{+}$ for
AdS BH in BI massive gravity with a non-abelian hair for negative
cosmological constant. Specific values of parameters are $~\nu=1,
~c=1, ~c_{1}=1, ~c_{2}=2$ and $b=1$}
\end{minipage}\hspace{3pc}%
\begin{minipage}{14pc}
\includegraphics[width=18pc]{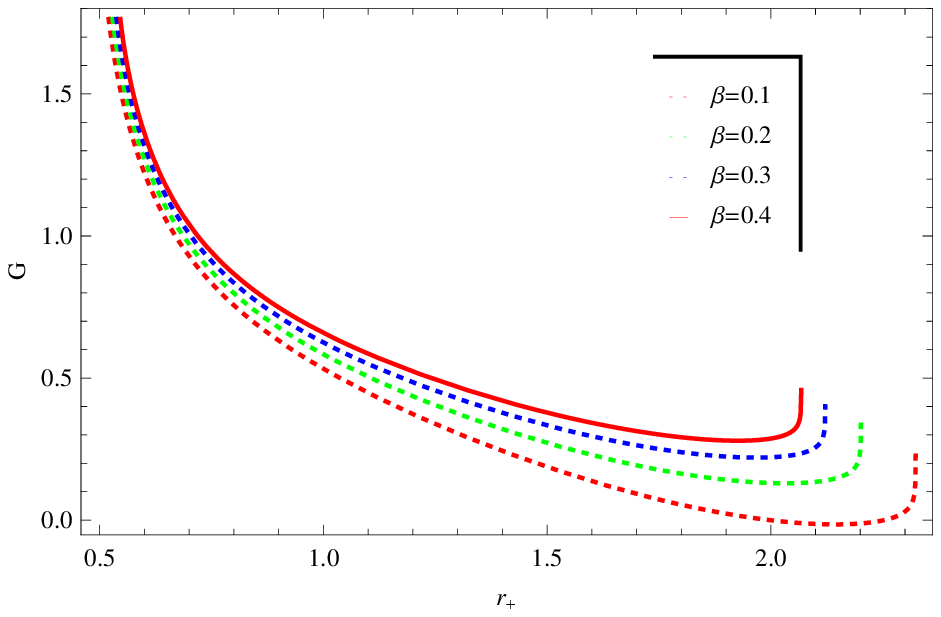}
\caption{\label{label}Plot of Gibbs free energy versus $r_{+}$ for
AdS BH in BI massive gravity with a non-abelian hair for positive
cosmological constant. Specific values of parameters are $ ~\nu=1,
~c=1, ~c_{1}=1, ~c_{2}=2$ and $b=1$}
\end{minipage}\hspace{3pc}%
\end{figure}
\begin{figure}
\begin{minipage}{14pc}
\includegraphics[width=18pc]{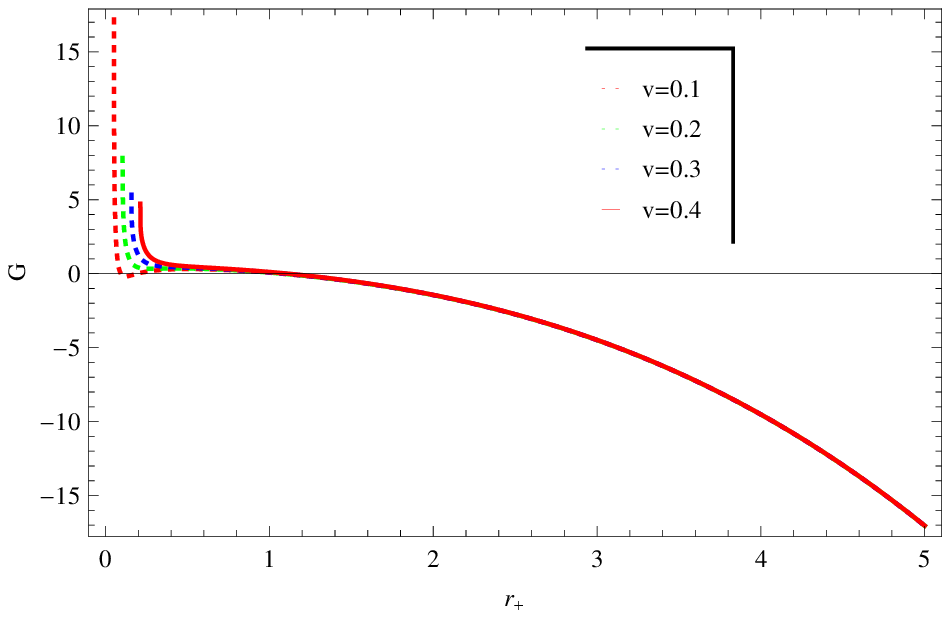}
\caption{\label{label}Plot of Gibbs free energy versus $r_{+}$ for
AdS BH in BI massive gravity with a non-abelian hair for negative
cosmological constant. Specific values of parameters are $\beta=1,
~c=1, ~c_{1}=1, ~c_{2}=2$ and $b=1$}
\end{minipage}\hspace{3pc}%
\begin{minipage}{14pc}
\includegraphics[width=18pc]{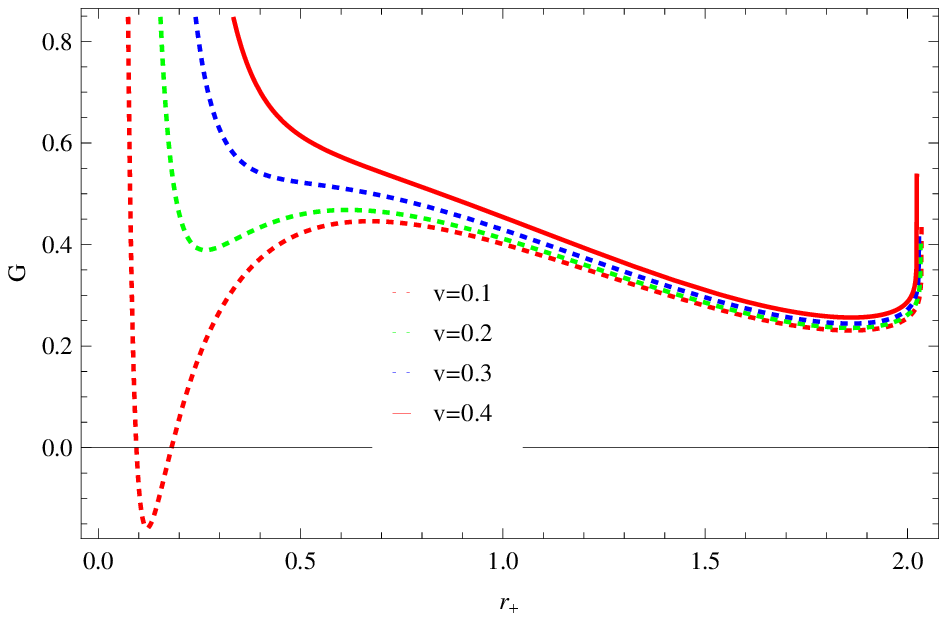}
\caption{\label{label}Plot of Gibbs free energy versus $r_{+}$ for
AdS BH in BI massive gravity with a non-abelian hair for positive
cosmological constant. Specific values of parameters are $\beta=1,
~c=1, ~c_{1}=1, ~c_{2}=2$ and $b=1$}
\end{minipage}\hspace{3pc}%
\end{figure}
In FIGs. \textbf{10} and \textbf{11}, we study the impact of
non-linear electrodynamics on the Gibbs free energy of AdS BH in BI
massive gravity with non-abelian hair. While in FIGS. \textbf{12}
and \textbf{13}, we demonstrate the impact of non-abelian hair on
the Gibbs free energy. In FIGs. \textbf{10} and \textbf{12},
non-linear electrodynamics and non-abelian hair effect slightly on
the Gibbs free energy. For very smaller values of horizon radius,
Gibbs free energy is positive while it becomes negative for
increasing values of $r_{+}$. The impact of non-linear
electrodynamics and non-abelian hair for the case of positive
cosmological constant in FIGs. \textbf{11} and FIGS. \textbf{13} are
remarkable because it makes the Gibbs energy positive. There is only
one case $\nu=0.1$ for which Gibbs energy is negative for very small
interval of horizon radius.
\begin{figure}
\begin{minipage}{14pc}
\includegraphics[width=18pc]{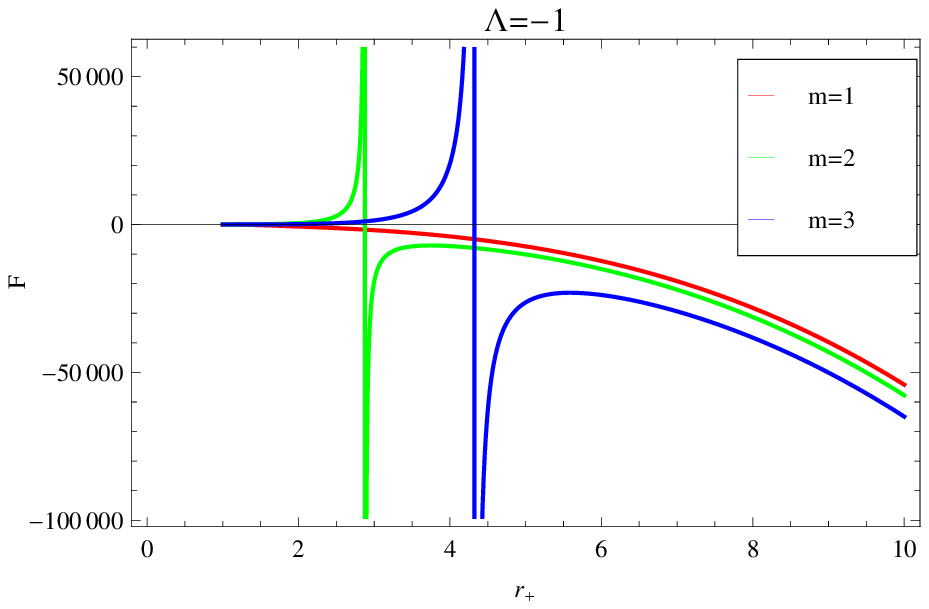}
\caption{\label{label}Plot of Helmholtz free energy versus $r_{+}$
for AdS BH in BI massive gravity with a non-abelian hair for
negative cosmological constant. Specific values of parameters are
$\beta=1, ~\nu=1, ~c=1, ~c_{1}=-1, ~c_{2}=2$ and $b=1$ }
\end{minipage}\hspace{3pc}%
\begin{minipage}{14pc}
\includegraphics[width=18pc]{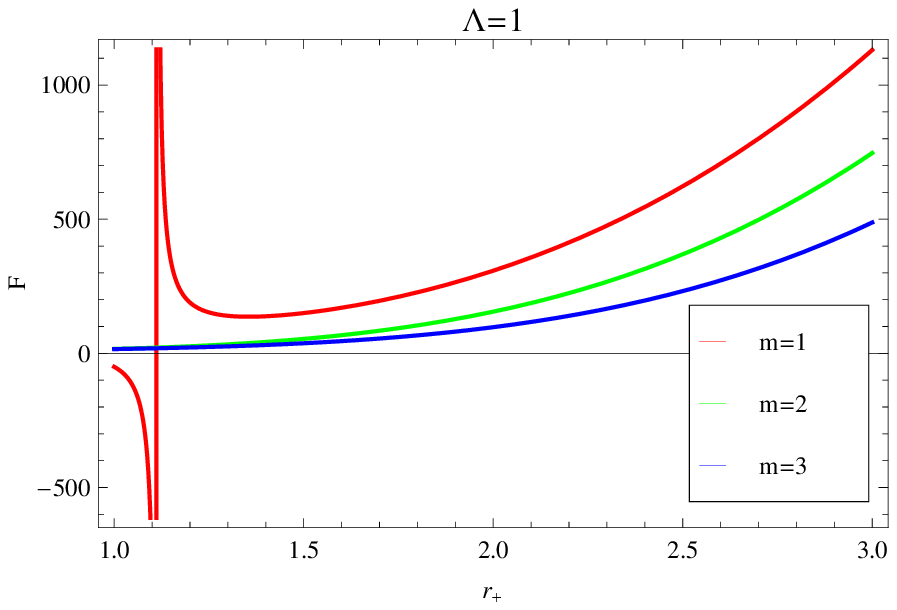}
\caption{\label{label}Plot of Helmholtz free energy versus $r_{+}$
for AdS BH in BI massive gravity with a non-abelian hair for
positive cosmological constant. Specific values of parameters are
$\beta=1, ~\nu=1, ~c=1, ~c_{1}=-1, ~c_{2}=2$ and $b=1$}
\end{minipage}\hspace{3pc}%
\end{figure}
\begin{figure}
\begin{minipage}{14pc}
\includegraphics[width=18pc]{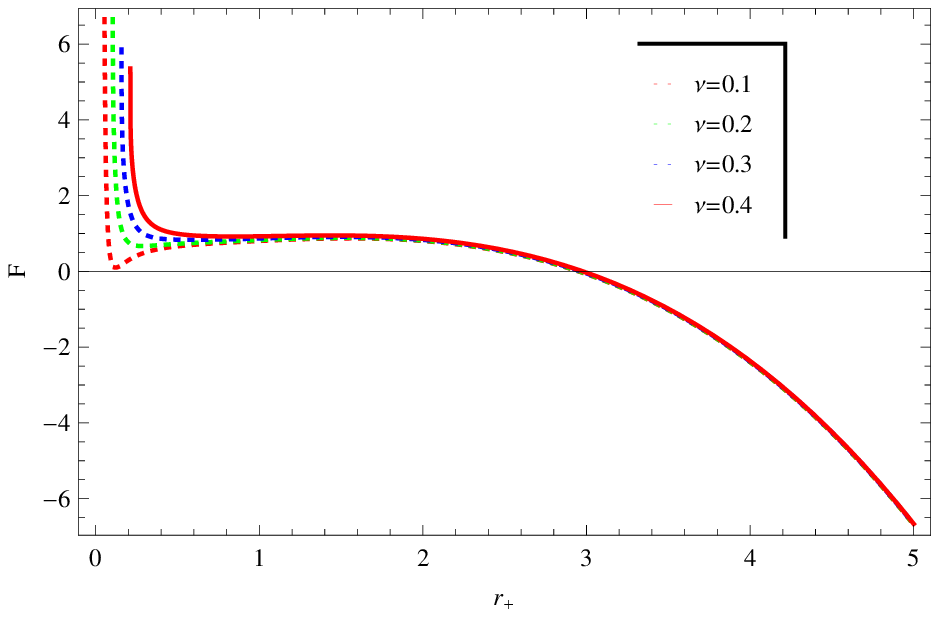}
\caption{\label{label}Plot of Helmholtz free energy versus $r_{+}$
for AdS BH in BI massive gravity with a non-abelian hair for
negative cosmological constant. Specific values of parameters are
$\beta=1, m=1, ~c=1, ~c_{1}=-1, ~c_{2}=2$ and $b=1$ }
\end{minipage}\hspace{3pc}%
\begin{minipage}{14pc}
\includegraphics[width=18pc]{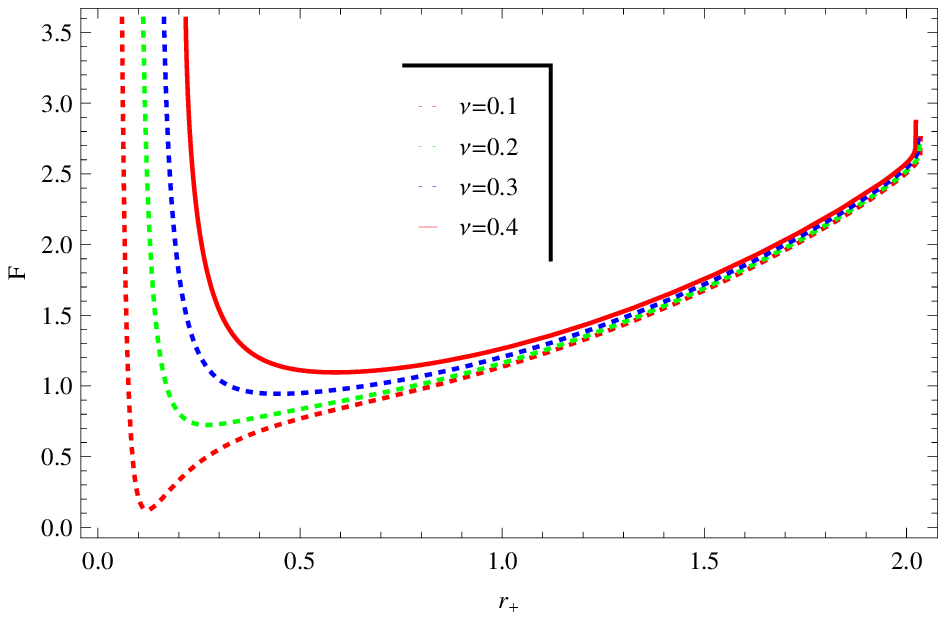}
\caption{\label{label}Plot of Helmholtz free energy versus $r_{+}$
for AdS BH in BI massive gravity with a non-abelian hair for
positive cosmological constant. Specific values of parameters are
$\beta=1, m=1, ~c=1, ~c_{1}=-1, ~c_{2}=2$ and $b=1$}
\end{minipage}\hspace{3pc}%
\end{figure}
Behavior of the Helmholtz free energy $F$ against the horizon radius
$r_{+}$ is represented in the FIGs.\textbf{14} and \textbf{15} for
both positive and negative cosmological constant at fixed values of
massless graviton. When $\Lambda=-1$, for $m=1$ the Helmholtz free
energy decreases and becomes negative for the larger horizon. When
$m=2$, the Helmholtz free energy is positive for SBH but becomes
negative when $r_{+}\geq2.8$. When $m=3$, the Helmholtz free energy
becomes negative for $r_{+}\geq4.3$. We conclude that the Helmholtz
free energy is negative for the larger horizon and higher values of
massless graviton for the negative cosmological constant. When
$\Lambda=1$, for $m=1$ the Helmholtz free energy is negative for
small horizon while for $r\geq1.1$, Helmholtz free energy becomes
positive. For $m=2$ and $m=3$, the Helmholtz free energy is positive
and increases for the larger values of horizon radius. We conclude
that, the Helmholtz free energy is higher for $m=1$ as compared with
$m=2$ and $m=3$. Also the Helmholtz free energy is positive, which
leads to stability for the larger horizon radius and positive
cosmological constant. FIGs. \textbf{16} and \textbf{17} demonstrate
Helmholtz free energy for positive and negative cosmological
constant at fixed values of $\nu$. For the case of negative
cosmological constant Helmholtz energy is positive for very small
horizon radius of all fixed values of $\nu$, then it turns to
negative for intermediate and large horizon. For positive
cosmological constant, Helmholtz energy is positive for all the
cases of $\nu$.

\section{Charged AdS BH With a Global Monopole}

The gravitational field of a global monopole which is an approximate
solution for the metric outside a monopole produce from the breaking
of a $O(3)$ symmetry is discussed by Barriola and Vilenkin
\cite{51}. When a charged AdS BHs swallows a global monopole the
general static spherically symmetric metric and gauged potential can
be written as \cite{52}
\begin{equation}\label{27}
d\tilde{s}^{2}=-\tilde f{\tilde{(r)}}d\tilde{t}^{2}+\frac{1}{\tilde
h{\tilde{(r)}}}+\tilde{r}^{2}(d\theta^{2}+sin^{2}\theta d\phi^{2}),
\end{equation}
\begin{equation}\label{28}
\tilde{A}=\frac{\tilde{q}}{\tilde{r}}d\tilde{t},~~\tilde
f{(\tilde{r})}=\tilde h{(\tilde{r})}=1-8\pi
\eta_{o}^{2}-\frac{2\tilde{m}}{\tilde{r}}+\frac{\tilde{q}^{2}}{\tilde{r}^{2}}+\frac{\tilde{r}^{2}}{l^{2}},
\end{equation}
where $\tilde{m}$ is the mass parameter and $\tilde{q}$ is an
electric charge parameter. Also, $l$ is associated with the
cosmological constant as $\Lambda=-\frac{3}{l^{2}}$. The coordinate
transformations are given by
\begin{equation}\label{29}
\tilde{t}=(1-8\pi \eta_{o}^{2})^{-\frac{1}{2}}t, \tilde{r}=(1-8\pi
\eta_{o}^{2})^{\frac{1}{2}}r.
\end{equation}
New parameters can be defined as follows
\begin{equation}\label{30}
m=(1-8\pi \eta_{o}^{2})^{-\frac{3}{2}}\tilde{m}, q=(1-8\pi
\eta_{o}^{2})^{-1}\tilde{q}, \eta^{2}=8 \pi\eta_{o}^{2},
\end{equation}
where $\eta$ is the symmetry breaking parameter. In view of Eqs.
(\ref{27}) and (\ref{29}), Eq. (\ref{30}) can be written as,
\begin{equation}\label{31}
ds^{2}=-f(r)dt^{2}+\frac{1}{h(r)}dr^{2}+(1-\eta^{2})r^{2}(d\theta^{2}+sin^{2}\theta
d\phi^{2}),
\end{equation}
the solution will take the form of the four-dimensional
Reissner-Nordst$\ddot{o}$m (RN) AdS BH if we set $\eta=0$, where
\begin{equation}\label{32}
A=\frac{q}{r}dt,
f(r)=h(r)=1-\frac{2m}{r}+\frac{q^{2}}{r^{2}}+\frac{r^{2}}{l^{2}},
\end{equation}
by using $l^{2}=-\frac{3}{\Lambda}$, we get
\begin{equation}\label{33}
f(r)=1-\frac{2m}{r}+\frac{q^{2}}{r^{2}}-\frac{\Lambda r^{2}}{3}.
\end{equation}
In order to calculate the thermodynamical quantities of charged AdS
BH with global monopole, we begin with the event horizon of the BH
situated at $r=r_{h}$, which is the largest root of the
$f(r)=h(r)=0$. At infinity, the electric charge $q$ along with the
potential can be evaluated by
\begin{equation}\label{34}
Q=\frac{1}{4\pi}\oint \textsl{F}^{\mu\omega}d^{2}\sum{\mu\omega}
=(1-\eta^{2})q,
\end{equation}
\begin{equation}\label{35}
\Phi=\xi^{\mu}A_{\mu}|_{r=r_{h}}=\frac{q}{r_{h}},
\end{equation}
where $\xi^{\mu}$ is the timelike killing vector
$(\partial_{t})^{\mu}$ \cite{53}. The Arnowitt-Deser-Misner (ADM)
mass $M$ of the system can be evaluated by the Komar integral
\begin{equation}\label{36}
M=\frac{1}{8\pi}\oint\xi_{(t)}^{\mu:\nu}d^{2}\sum_{\mu\nu},
\end{equation}
\begin{equation}\label{37}
M=(1-\eta^{2})m.
\end{equation}
In order to find the expression for mass put $f(r)=0$ in Eq.
(\ref{33}), we obtain
\begin{equation}\label{38}
m=\frac{r}{2}+\frac{q^{2}}{2r}-\frac{\Lambda r^{3}}{6}.
\end{equation}
By using Eqs. (\ref{34}), (\ref{37}) and (\ref{38}), we get
\begin{equation}\label{39}
M=(1-\eta^{2})\bigg(\frac{r_{+}}{2}+\frac{Q^{2}}{(1-\eta^{2})^{2}2r_{+}}-\frac{\Lambda
r_{+}^{3}}{6}\bigg).
\end{equation}

The Hawking temperature $T$ of the charged AdS BH with global
monopole is given by
\begin{equation}\label{40}
T=\frac{f'(r)}{4\pi}=\frac{1}{4\pi r_{+}}\bigg(1-\Lambda
r_{+}^{2}-\frac{Q^{2}}{(1-\eta^{2})^{2}r_{+}^{2}}\bigg).
\end{equation}
The area $A$ at the horizon of the charged AdS BH with golobal
monopole can be obtained as
\begin{equation}\label{41}
A=\int_{r=r_{+}}\sqrt{g_{\theta\theta}g_{\phi\phi}}d\theta
d\phi=4\pi(1-\eta^{2})r_{+}^{2}.
\end{equation}
The entropy $S_{o}$ can be evaluated by using the above equation is
given by
\begin{equation}\label{42}
S_{o}=\frac{A}{4}=\pi(1-\eta^{2})r_{+}^{2}.
\end{equation}
The logarithmic correction term for the entropy of charged AdS BH
with global monopole can be obtained by using Eq. (\ref{11})
\begin{equation}\label{43}
S=\pi(1-\eta^{2})r_{+}^{2}-\frac{b}{2}\log\bigg[\frac{(1-\eta^{2})\bigg(1
+\frac{Q^{2}}{r_{+}^{2}(-1+\eta^{2})^{2}}-r_{+}^{2}\Lambda\bigg)^{2}}{16\pi}\bigg].
\end{equation}
The thermodynamic volume $V=(\frac{\partial M}{\partial P})$ can be
given as
\begin{equation}\label{45}
V=\frac{4\pi(1-\eta^{2})r_{+}^{3}}{3}.
\end{equation}
From the above results, it is clear that the thermodynamical
quantities such as, the electric charge $Q$, the ADM mass $M$, the
area at the horizon of the BH $A$, the thermodynamic volume $V$, the
corrected term of the entropy $S$ and the temperature $T$ are
associated with the the symmetry breaking parameter $\eta$. The
first law of thermodynamics holds for the above thermodynamical
quantities for details see ref \cite{a100}. We step forward to
analyze the thermal stability, phase transition, local and global
stability in the presence of the grand canonical ensemble and
canonical ensemble. The equation of state for charged AdS BH with
global monopole take the following form
\begin{equation}\label{49}
P=\frac{Q^2+\left(-1+\eta^2\right)^2 r^2 (-1+4 \pi  r T)}{8
\left(-1+\eta^2\right)^2 \pi  r^4}.
\end{equation}

\begin{figure} \centering
\epsfig{file=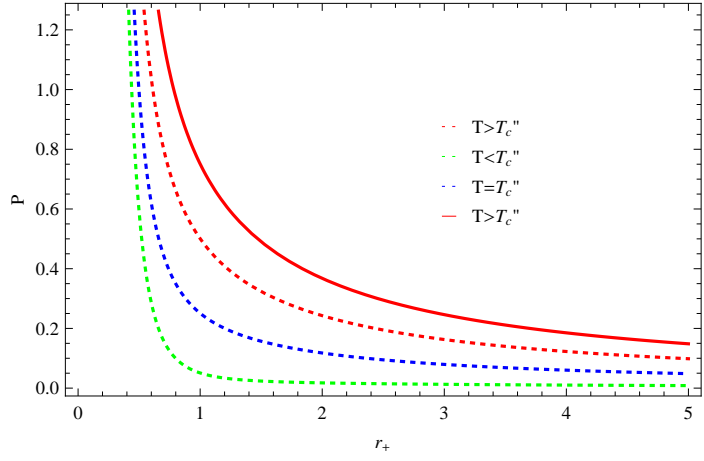,width=.60\linewidth}\caption{Plot of $P$ versus
$r_{+}$ for charged AdS BH with a global monopole.}
\end{figure}
FIG. \textbf{18} demonstrates $P$ versus $r_{+}$ for charged AdS BH
with a global monopole. The two upper red lines correspond to the
``ideal gas" phase for $T> T_{c}$, the critical isotherm $T = T_{c}$
is denoted by the dotted blue line, lower dotted green correspond to
temperatures smaller than the critical temperature.

Further, to analyze and study the phase transition of the charged
AdS BH with a global monopole, we move to evaluate the important
thermodynamical quantities such as the heat capacities $C_{p}$ and
$C_{v}$. Generally, when the heat capacity is positive it implies
that the BH is stable and negative heat capacity implies that BH is
unstable. Instability of a BH means that the BH cannot bear even a
small perturbation and proceed to disappear. Heat capacity for the
mentioned BH can be evaluated by the using the relation Eq.
(\ref{17}) as
\begin{eqnarray}\label{50}
C_{v}&=&\frac{1}{3Q^{2}-r_{+}^{2}(-1+\eta^{2})^{2}(1+r_{+}^{2}\Lambda)}\times
\bigg(r_{+}^{3}(-1+\eta^{2})^{2}\big(1-\frac{Q^{2}}{r_{+}^{2}(-1+\eta^{2})^{2}}-r_{+}^{2}\Lambda\big)\\\nonumber&\times&\big(-2\pi
r_{+}(-1+\eta^{2})+\frac{2b(Q^{2}+r_{+}^{4}(-1+\eta^{2})^{2}\Lambda)}{r_{+}(Q^{2}-r^{2}(-1+\eta^{2})^{2}(-1+r_{+}^{2}\Lambda))}\big)\bigg).
\end{eqnarray}
\begin{figure}
\begin{minipage}{14pc}
\includegraphics[width=18pc]{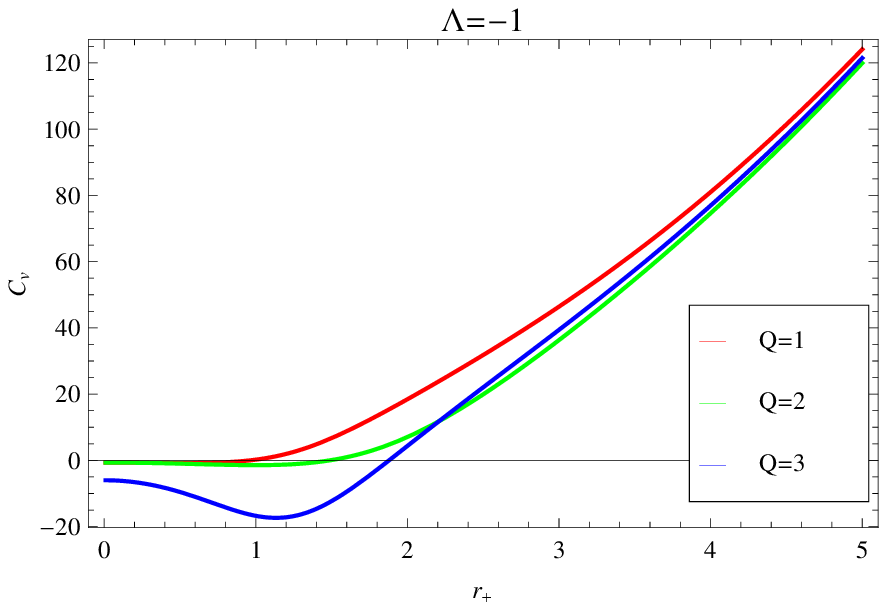}
\caption{\label{label}Plot of $C_{v}$ versus $r{+}$ for charged AdS
BH with a global monopole for negative cosmological constant.}
\end{minipage}\hspace{3pc}%
\begin{minipage}{14pc}
\includegraphics[width=18pc]{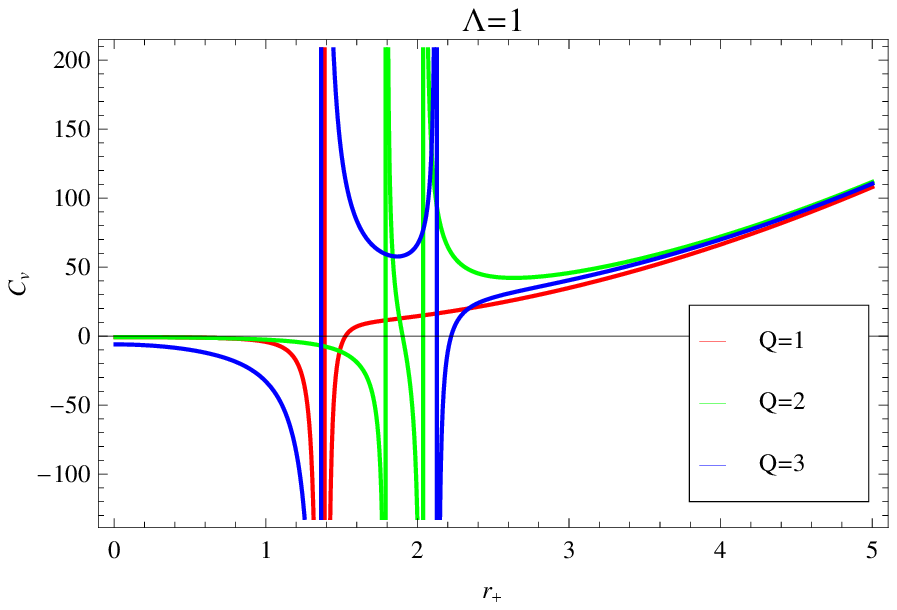}
\caption{\label{label}Plot of $C_{v}$ versus $r_{+}$ for charged AdS
BH with a global monopole for positive cosmological constant.}
\end{minipage}\hspace{3pc}%
\end{figure}
\begin{figure}
\begin{minipage}{14pc}
\includegraphics[width=18pc]{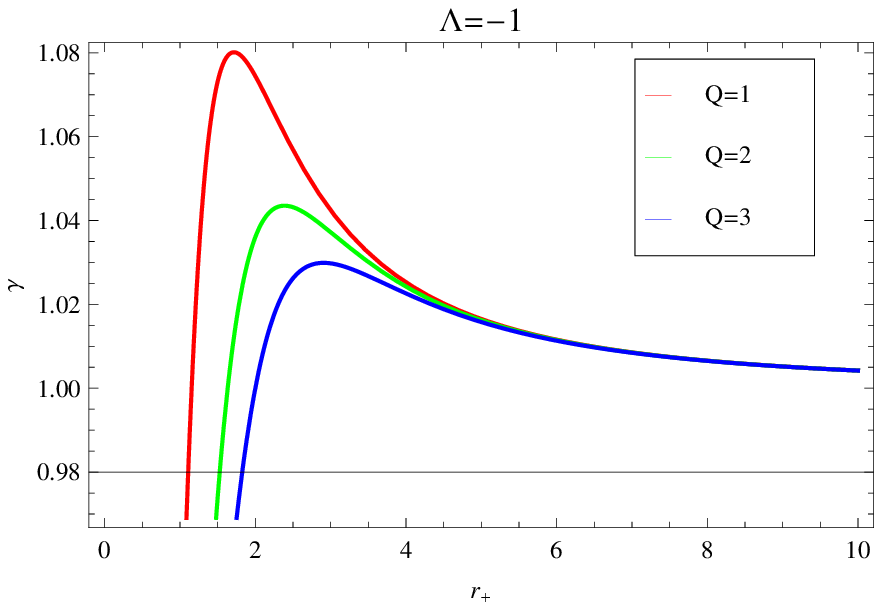}
\caption{\label{label}Plot of $\gamma$ versus $r_{+}$ for charged
AdS BH with a global monopole for negative cosmological constant.}
\end{minipage}\hspace{3pc}%
\begin{minipage}{14pc}
\includegraphics[width=18pc]{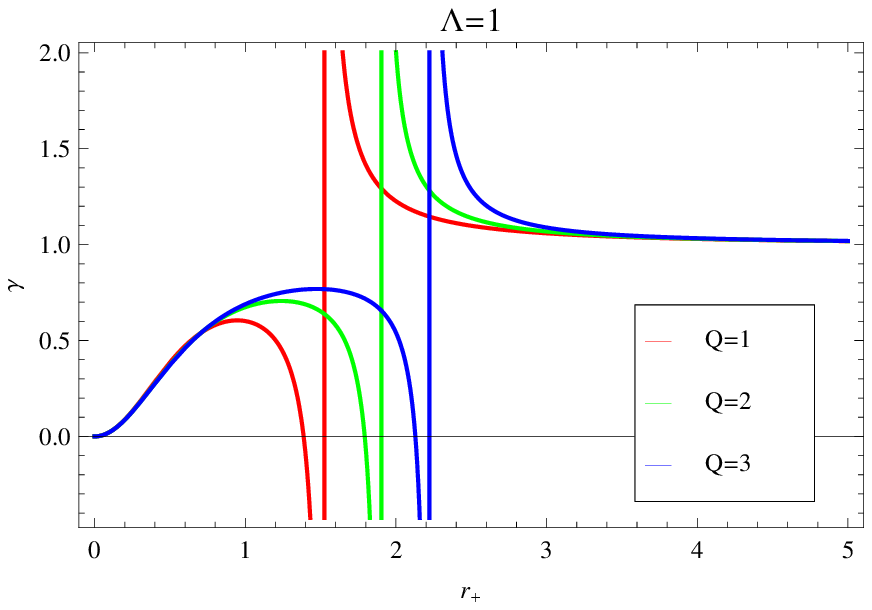}
\caption{\label{label}Plot of $\gamma$ versus $r_{+}$ for charged
AdS BH with a global monopole for positive cosmological constant.}
\end{minipage}\hspace{3pc}%
\end{figure}
The graphical analysis of heat capacity at constant volume is
represented in FIGs. \textbf{19} and \textbf{20} for negative and
positive cosmological constant respectively. In Fig. \textbf{19},
when $\Lambda=-1$, the $C_{v}$ is negative for $0\leq r_{+}\leq0.1,
~0\leq r_{+}\leq1.5$ and $0\leq r_{+}\leq1.89$ for $Q=1, ~Q=2$ and
$Q=3$, respectively. The heat capacity is positive for higher values
of horizon radius. We conclude that, for negative cosmological
constant, phase transition takes place and BH is unstable for lower
values of horizon radius but stable for larger values of horizon
radius. In FIG. \textbf{20}, when $Q=1$, the curve has two regions
with one divergent point. The small radius region has negative heat
capacity therefore BH is unstable in this region while the large
radius region has positive heat capacity therefore it is stable.
When $Q=2$, the curve has four regions with three divergent points.
For the small horizon, the heat capacity is negative and thus it is
unstable. For the first part of the intermediate part of the
horizon, the heat capacity is positive and BH is stable in this
region. But in the second part of the intermediate BH (IBH) the heat
capacity is negative and so BH is unstable. The LBH has positive
heat capacity and therefore BH is stable in this region. The phase
transition takes place in between SBH first and second part of IBH
and LBH. When $Q=3$, the curve has two divergent points in three
regions. The phase transition takes place again as shown in the
figure.

Now, the heat capacity when the heat is added to the system at the
constant pressure $C_{P}$ can be evaluated by the using the relation
Eq.(\ref{19}) as
\begin{equation}\label{51}
C_{p}=-\frac{2\pi
r_{+}^{2}(-1+\eta^{2})(Q^{2}+r_{+}^{2}(-1+\eta^{2})^{2}(-1+r_{+}^{2}\Lambda))}{-3Q^{2}+r_{+}^{2}(-1+\eta^{2})^{2}(1+r_{+}^{2}\Lambda)}
\end{equation}
Also, $\gamma$ turns out to be
\begin{eqnarray}\label{52}
\gamma&=&\bigg(\pi
r_{+}^{2}(-1+\eta^{2})(-Q^{2}+r_{+}^{2}(-1+\eta^{2})^{2}(-1+r^{2}\Lambda))\bigg)\bigg(b(Q^{2}+r_{+}^{4}(-1+\eta^{2})^{2}\Lambda)\\\nonumber&-&\pi
r_{+}^{2}(-1+\eta^{2})(Q^{2}+r_{+}^{2}(-1+\eta^{2})^{2}(1-r_{+}^{2}\Lambda))\bigg)^{-1}.
\end{eqnarray}
FIGs. \textbf{21} and \textbf{22} represent the graphical analysis
on $\gamma$ versus $r_{+}$ for negative and positive cosmological
constant, respectively. In FIG. \textbf{21} for $\Lambda=-1$,
$\gamma$ is higher for the lower values of $Q$. For $Q=1, ~Q=2$ and
$Q=3$, the values of $\gamma$ are highest at $r_{+}=1.6, ~r_{+}=2.5$
and $r_{+}=3$, respectively. For $r_{+}\geq5$, the value of $\gamma$
starts approaching to the same value for all different values of
$Q$. We conclude that for the negative cosmological constant and for
all different values of $Q$, the values of $\gamma$ are positive. In
FIG. \textbf{22}, we analyze the graph for three different values of
$Q$ and discuss the case for $\Lambda=1$. When $Q=1$, for the small
radial region, $\gamma$ is positive for the first part of the
intermediate region then $\gamma$ becomes negative and again for the
second part of intermediate region for large radial region, $\gamma$
becomes positive. When $Q=1$ and $Q=2$, both curves show similar
behavior as per $Q=1$. Hence, we conclude that the for the smaller
values of radius, the $\gamma$ is negative while for the larger
values of the radius, the $\gamma$ is positive and the curve
approaches to the same value of $\gamma$.

Now, we investigate the global stability of the charged AdS BH with
a global monopole by using the Gibbs free energy or free energy in
the grand canonical ensemble. The Gibbs free energy can be derived
by using Eq.(\ref{23}). The expression for the Gibbs free energy of
the charged AdS BH with global monopole is
\begin{eqnarray}\label{53}
G&=&\frac{1}{24(-1+\eta^2)^2 r^3}(12 (-1+\eta^2) r^2
(Q^2+(-1+\eta^2) r(2 M+(-1+\eta^2)
r))+4(-1\\\nonumber&+&\eta^2)r^2(-3 Q^2+(-1+\eta^2)^2 r^2(-3+\Lambda
r^2))-(3(Q^2+(-1+\eta^2)^2 r^2(-1+\Lambda
r^2))\\\nonumber&\times&(2(-1+\eta^2) \pi r^2+b
\text{Log}[\frac{(1-\eta^2)(1+\frac{Q^2}{(-1+\eta^2)^2 r^2}-\Lambda
r^2)^2}{16 \pi }]))(\pi)^{-1})
\end{eqnarray}
FIGs. \textbf{23} and \textbf{24} depict the graph between the Gibbs
free energy $G$ against the radius of the BH at the horizon $r_{+}$
at different values of $Q$. In FIG. \textbf{23}, when $\Lambda=-1$,
apparently there are no divergent points. In the small region
radius, the Gibbs free energy is negative for all different values
of $Q$. But the Gibbs free energy is positive and displays a smooth
curve in the IBH and again its negative in the LBH. Thus, we
conclude that the charged AdS BH with a global monopole is unstable
in the SBH and LBH but, its stable in the IBH. In FIG. \textbf{24},
for all different values of $Q$, the Gibbs free energy is negative
for the SBH and positive for the LBH. When $Q=4$, the Gibbs free
energy is highest as compared to for the $Q=3$, $Q=2$ and $Q=1$.
Thus, the Gibbs free energy is higher for the higher the value of
$Q$. We conclude that the charged AdS BH with a global monopole is
globally stable for the larger horizon radius.

Now, for investing the thermodynamically global stability in the
canonical ensemble, we will consider the charged AdS BH with a
global monopole in the canonical ensemble. The free energy or the
Helmholtz free energy for charged AdS BH with a global monopole can
be derived by using the Eq. (\ref{25}) as
\begin{eqnarray}\nonumber
F&=&\frac{1}{24}\bigg(\frac{4(-1+\eta^{2})(-3(q^{2}+r_{+}^{2})+r_{+}^{4}\Lambda)}{r_{+}}-\frac{1}{\pi}\bigg(2(1-\frac{Q^{2}}{r_{+}^{2}(-1+\eta^{2})^{2}}-r_{+}^{2}\Lambda)(-2\pi
r_{+}(-1\\\nonumber&+&\eta^{2})+\frac{2b(Q^{2}+r_{+}^{4}(-1+\eta^{2})^{2}\Lambda)}{r_{+}(Q^{2}-r_{+}^{2}(-1+\eta^{2})^{2}(-1+r_{+}^{2}\Lambda))})\bigg)-\frac{1}{\pi
r_{+}^{3}(-1+\eta^{2})^{2}}\bigg(3(Q^{2}+r_{+}^{2}(-1+\eta^{2})^{2}\\\nonumber&\times&(-1+r_{+}^{2}\Lambda))(2\pi
r_{+}^{2}(-1+\eta^{2})\\\label{54}&+&b\log[\frac{(1-\eta^{2})(-1+\frac{Q^{2}}{r_{+}^{2}(-1+\eta^{2})^{2}}+r_{+}^{2}\Lambda)}{16\pi}])\bigg)\bigg).
\end{eqnarray}
\begin{figure}
\begin{minipage}{14pc}
\includegraphics[width=18pc]{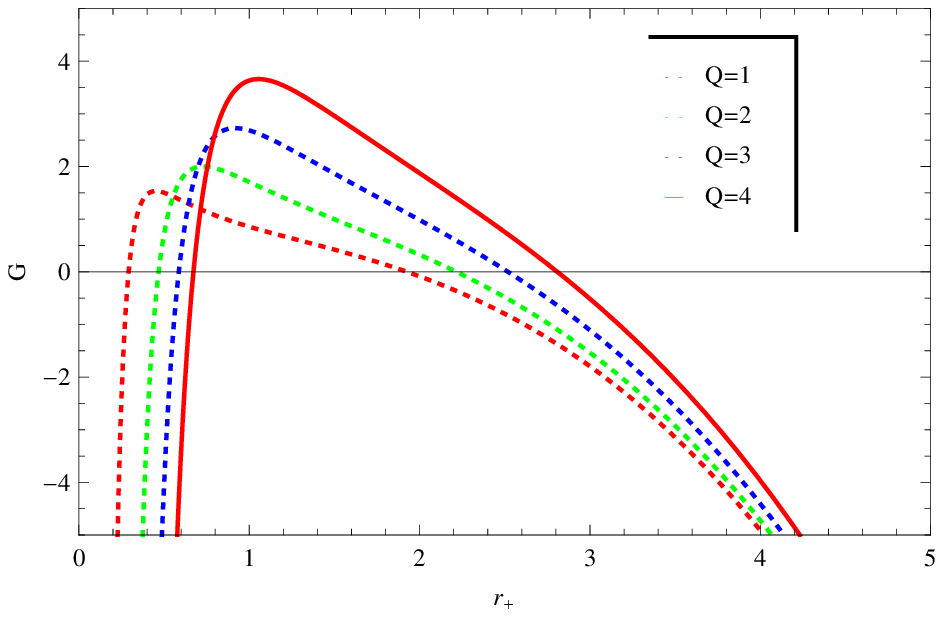}
\caption{\label{label}Plot of $G$ versus $r_{+}$ for charged AdS BH
with a global monopole for negative cosmological constant.}
\end{minipage}\hspace{3pc}%
\begin{minipage}{14pc}
\includegraphics[width=18pc]{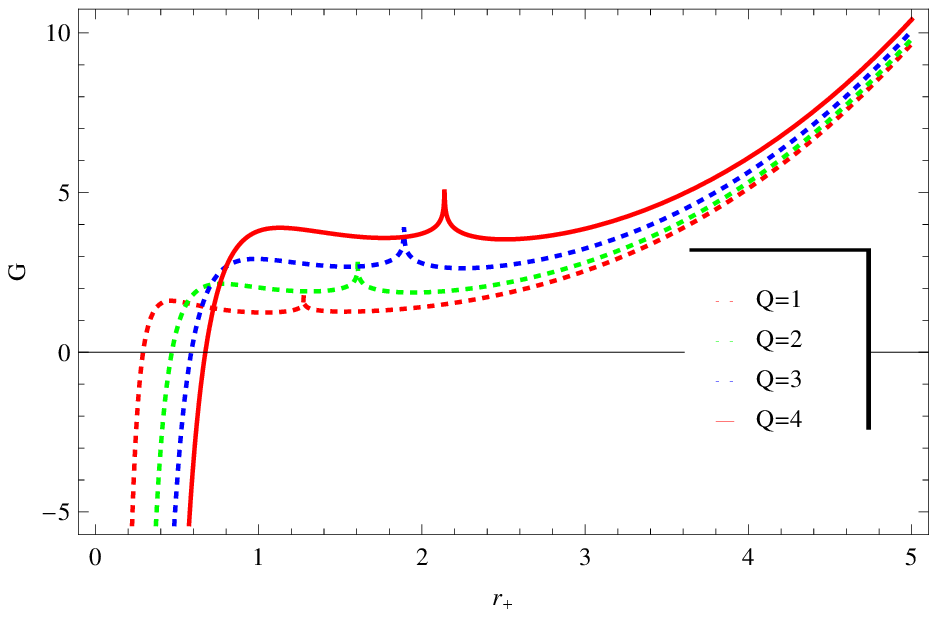}
\caption{\label{label}Plot of $G$ versus $r_{+}$ for charged AdS BH
with a global monopole for positive cosmological constant.}
\end{minipage}\hspace{3pc}%
\end{figure}
\begin{figure}
\begin{minipage}{14pc}
\includegraphics[width=18pc]{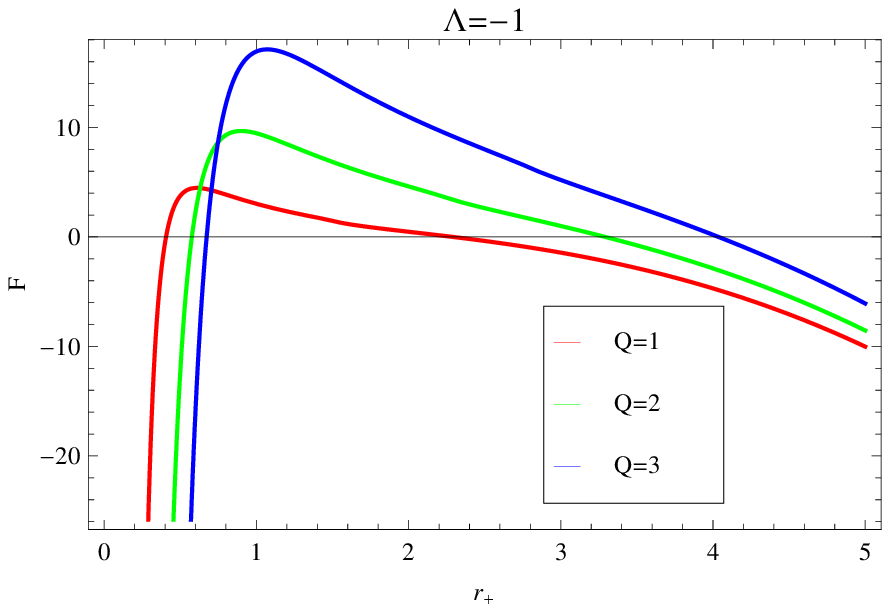}
\caption{\label{label}Plot of $F$ versus $r_{+}$ for charged AdS BH
with a global monopole for negative cosmological constant.}
\end{minipage}\hspace{3pc}%
\begin{minipage}{14pc}
\includegraphics[width=18pc]{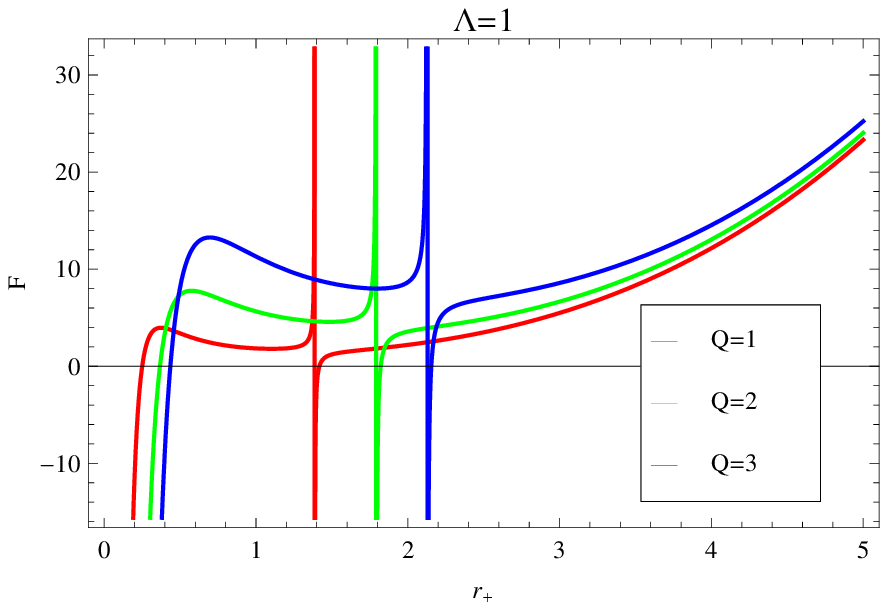}
\caption{\label{label}Plot of $F$ versus $r_{+}$ for charged AdS BH
with a global monopole for positive cosmological constant.}
\end{minipage}\hspace{3pc}%
\end{figure}
The graphical analysis of Helmholtz free energy for charged AdS BH
with a global monopole for specific values of different parameters
is represented  in the FIGs. \textbf{25} and \textbf{26}. In FIG.
\textbf{25}, for $Q=1, ~Q=2$ and $Q=3$, when the cosmological
constant is negative the Helmholtz free energy is positive for
$0.25\leq r_{+}\leq1.5, ~0.6\leq r_{+}\leq2.25$ and $0.65\leq
r_{+}\leq2.8$, respectively. The Helmholtz free energy is negative
for the larger horizon radius thus the charged AdS BH with a global
monopole is unstable for LBH. In FIG. \textbf{26}, for $\Lambda=1$,
Helmholtz free energy displays an interesting and yet complicated
behavior. For $Q=1$, the Helmholtz free energy is negative for the
SBH but it is positive for the IBH and LBH. Similarly, for $Q=2$ and
$Q=3$, the Helmholtz free energy is negative for the SBH and
positive for the IBH and LBH. The Helmholtz free energy is
increasing and stays positive for the larger horizon radius. The
Helmholtz free energy is higher for the higher values of $Q$. Thus,
we conclude that for the lower horizon radius the charged AdS BH
with a global monopole is globally unstable and for the higher
horizon radius it is globally stable.

Our purpose was to check that does presence of monopole has any
influence on the phase behavior of the charged AdS BH with global
monopole. The above calculations and the graphical representation
implies that the charged AdS BH with a global monopole undergo a
small to large BH phase transition which is remarkably explained
through the graphical representation in \textbf{FIGs. 19-26}. Until
now, we have analyzed and discussed the local thermodynamical
properties of the charged AdS BH with a global monopole.  We have
find that the charged AdS BH with a global monopole experiences a
SBH and LBH phase transition, which is interestingly remarkable. If
we compare our results with RN AdS BH, we can easily see that the
existence of the global monopole has influence on the critical
points but the law of corresponding states remain unchanged and
constant as stated above.

\subsection{Tables} Table \textbf{1}: Summary table for the heat
capacity $C_{v}$ versus
$r_{+}$ for $\Lambda=-1$ and $\eta=0.5$. \\\\

\begin{tabular}{|c|c|c|c|}
\hline $C_{v}$ & $Q$ & Horizon radius $r_{+}$ & Stability \\
\hline
& Q=1 & $0\leq r_{+}\leq0.1$ & unstable \\
negative & $Q=2$ & $0\leq r_{+}\leq1.5$ & unstable \\
& $Q=3$ & $0\leq r_{+}\leq1.89$ & unstable \\ \hline
& $Q=1$ & $r_{+}=0.11$ & Phase Transition \\
$C_{v}=0$ & $Q=2$ & $r_{+}=0.51$ & Phase Transition \\
& $Q=3$ & $r_{+}=0.91$ & Phase Transition \\ \hline
& $Q=1$ & $r_{+}\geq0.12$ & stable \\
positive & $Q=2$ & $r_{+}\geq0.52$ & stable \\
& $Q=3$ & $r_{+}\geq0.2$ & stable \\ \hline
\end{tabular}\\\\

Table \textbf{2}: Summary table for the heat capacity $C_{v}$ versus
$r_{+}$ for $\Lambda=1$ and $\eta=0.5$.\\\\

\begin{tabular}{|c|c|c|c|}
\hline $C_{v}$ & $Q$ & Horizon radius $r_{+}$ & Stability \\
\hline
& Q=1 & $0\leq r_{+}\leq1.26, ~1.41\leq r_{+}\leq1.44$ & unstable \\
negative & $Q=2$ & $0\leq r_{+}\leq1.28, ~2.15\leq r_{+}\leq2.19$ & unstable \\
& $Q=3$ & $0\leq r_{+}\leq1.79, ~1.86\leq r_{+}\leq2.19$ & unstable \\
\hline
& $Q=1$ & $r_{+}=0.14, 1.5$ & Phase Transition \\
$C_{v}=0$ & $Q=2$ & $r_{+}=1.29, 2.16, 2.10$ & Phase Transition \\
& $Q=3$ & $r_{+}=1.8, 1.85, 2.2$ & Phase Transition \\ \hline
& $Q=1$ & $r_{+}\geq1.51$ & stable \\
positive & $Q=2$ & $1.3\leq r_{+}\leq 2.15, ~r_{+}\geq2.11$ & stable \\
& $Q=3$ & $1.81\leq r_{+}\leq 1.84, ~r_{+}\geq2.21$ & stable \\
\hline
\end{tabular}\\\\

Table \textbf{3}: Summary table for the $\gamma$ versus $r_{+}$ for
$\Lambda=-1$ and $\eta=0.5$. \\\\

\begin{tabular}{|c|c|c|c|}
\hline $\gamma$ & $Q$ & Horizon radius $r_{+}$ & Stability \\
\hline
& Q=1 & $0.51\leq r_{+}\leq0.52$ & unstable \\
negative & $Q=2$ & $0.69\leq r_{+}\leq0.71$ & unstable \\
& $Q=3$ & $0.85\leq r_{+}\leq0.89$ & unstable \\ \hline
& $Q=1$ & $r_{+}=0.53$ & Phase Transition \\
$\gamma=0$ & $Q=2$ & $r_{+}=0.72$ & Phase Transition \\
& $Q=3$ & $r_{+}=0.9$ & Phase Transition \\ \hline
& $Q=1$ & $r_{+}\geq0.54$ & stable \\
positive & $Q=2$ & $r_{+}\geq0.73$ & stable \\
& $Q=3$ & $r_{+}\geq0.91$ & stable \\ \hline
\end{tabular}\\\\

Table \textbf{4}: Summary table for the $\gamma$ versus $r_{+}$ for
$\Lambda=1$ and $\eta=0.5$. \\\\

\begin{tabular}{|c|c|c|c|}
\hline $\gamma$ & $Q$ & Horizon radius $r_{+}$ & Stability \\
\hline
& Q=1 & $1.41\leq r_{+}\leq1.5$ & unstable \\
negative & $Q=2$ & $1.81\leq r_{+}\leq1.9$ & unstable \\
& $Q=3$ & $2.18\leq r_{+}\leq2.21$ & unstable \\ \hline
& $Q=1$ & $r_{+}=1.40, 1.51$ & Phase Transition \\
$\gamma=0$ & $Q=2$ & $r_{+}=1.80, 1.2$ & Phase Transition \\
& $Q=3$ & $r_{+}=2.17, 2.22$ & Phase Transition \\ \hline
& $Q=1$ & $0\leq r_{+}\leq1.39, ~r_{+}\geq1.52$ & stable \\
positive & $Q=2$ & $0\leq r_{+}\leq1.79, ~r_{+}\geq1.21$ & stable \\
& $Q=3$ & $0\leq r_{+}\leq2.16, ~r_{+}\geq2.23$ & stable \\ \hline
\end{tabular}\\\\

Table \textbf{5}: Summary table for the $G$ versus $r_{+}$ for
$\Lambda=-1$ and $\eta=0.5$. \\\\

\begin{tabular}{|c|c|c|c|}
\hline $G$ & $Q$ & Horizon radius $r_{+}$ & Stability \\
\hline
& Q=1 & $0.21\leq r_{+}\leq0.24, ~r_{+}\geq2$ & unstable \\
negative & $Q=2$ & $0.36\leq r_{+}\leq0.4, ~r_{+}\geq2.23$ & unstable \\
& $Q=3$ & $0.41\leq r_{+}\leq0.42, ~r_{+}\geq2.6$  & unstable \\
\hline
& $Q=1$ & $r_{+}=0.23, 2$ & Phase Transition \\
$G=0$ & $Q=2$ & $r_{+}=0.45, 2.23$ & Phase Transition \\
& $Q=3$ & $r_{+}=0.48, 2.59$ & Phase Transition \\ \hline
& $Q=1$ & $0.23\leq r_{+}\leq2$ & stable \\
positive & $Q=2$ & $0.45\leq r_{+}\leq2.23$ & stable \\
& $Q=3$ & $0.48\leq r_{+}\leq2.59$ & stable \\ \hline
\end{tabular}\\\\

Table \textbf{6}: Summary table for the $G$ versus $r_{+}$ for
$\Lambda=1$ and $\eta=0.5$. \\\\

\begin{tabular}{|c|c|c|c|}
\hline $G$ & $Q$ & Horizon radius $r_{+}$ & Stability \\
\hline
& Q=1 & $0.22\leq r_{+}\leq0.24$ & unstable \\
negative & $Q=2$ & $0.37\leq r_{+}\leq0.39$ & unstable \\
& $Q=3$ & $0.43\leq r_{+}\leq0.45$ & unstable \\ \hline
& $Q=1$ & $r_{+}=0.25$ & Phase Transition \\
$G=0$ & $Q=2$ & $r_{+}=0.40$ & Phase Transition \\
& $Q=3$ & $r_{+}=0.46$ & Phase Transition \\ \hline
& $Q=1$ & $r_{+}\geq0.26$ & stable \\
positive & $Q=2$ & $r_{+}\geq0.41$ & stable \\
& $Q=3$ & $r_{+}\geq0.47$ & stable \\ \hline
\end{tabular}\\\\

Table \textbf{7}: Summary table for the $F$ versus $r_{+}$ for
$\Lambda=-1$ and $\eta=0.5$. \\\\

\begin{tabular}{|c|c|c|c|}
\hline $F$ & $Q$ & $r_{+}$ & Stability \\
\hline
& Q=1 & $0.23\leq r_{+}\leq0.30, ~r_{+}\geq2.4$ & unstable \\
negative & $Q=2$ & $0.42\leq r_{+}\leq0.49, ~r_{+}\geq3.4$ & unstable \\
& $Q=3$ & $0.55\leq r_{+}\leq0.60, ~r_{+}\geq4.1$  & unstable \\
\hline
& $Q=1$ & $r_{+}=0.31, 2.39$ & Phase Transition \\
$F=0$ & $Q=2$ & $r_{+}=0.50, 3.39$ & Phase Transition \\
& $Q=3$ & $r_{+}=0.61, 4.09$ & Phase Transition \\ \hline
& $Q=1$ & $0.32\leq r_{+}\leq2.38$ & stable \\
positive & $Q=2$ & $0.51\leq r_{+}\leq3.38$ & stable \\
& $Q=3$ & $0.62\leq r_{+}\leq4.08$ & stable \\ \hline
\end{tabular}\\\\

Table \textbf{8}: Summary table for the $F$ versus $r_{+}$ for
$\Lambda=1$ and $\eta=0.5$. \\\\

\begin{tabular}{|c|c|c|c|}
\hline $F$ & $Q$ & Horizon radius $r_{+}$ & Stability \\
\hline
& Q=1 & $0.19\leq r_{+}\leq0.22, ~1.39\leq r_{+}\leq1.40$ & unstable \\
negative & $Q=2$ & $0.3\leq r_{+}\leq0.38, ~1.79\leq r_{+}\leq1.80$ & unstable \\
& $Q=3$ & $0.4\leq r_{+}\leq0.42, ~2.1\leq r_{+}\leq2.2$ & unstable
\\ \hline
& $Q=1$ & $r_{+}=0.23, 1.38, 1.41$ & Phase Transition \\
$F=0$ & $Q=2$ & $r_{+}=0.39, 1.78, 1.81$ & Phase Transition \\
& $Q=3$ & $r_{+}=0.43, 2.09, 2.21$ & Phase Transition \\ \hline
& $Q=1$ & $0.23\leq r_{+}\leq1.37, ~r_{+}\geq1.42$ & stable \\
positive & $Q=2$ & $0.39\leq r_{+}\leq1.77, ~r_{+}\geq1.82$ & stable \\
& $Q=3$ & $0.43\leq r_{+}\leq2.08, ~r_{+}\geq2.22$ & stable \\
\hline
\end{tabular}\\\\
\section{Conclusions}

In this paper, we have analyzed the effects of thermal fluctuations
on the AdS BH in BI massive gravity with a non-abelian hair and the
charged AdS BH with a global monopole. Utilizing the logarithmic
correction of the entropy, we have calculated the conserved and
thermodynamical quantities for both BHs. We have studied the
$P-r_{+}$ behavior and local stability for the AdS BH in BI massive
gravity with a non-abelian hair through heat capacity. Moreover, we
have discussed impact of parameter $\nu$ on the local stability of
BH. We have investigated the behavior of $\gamma$ for negative and
positive cosmological constant. We have studied the impact of the
different parameters such as, $\beta$ and $\nu$ on the Gibbs free
energy. We also have analyzed the impact of non-abelian hair $\nu$
on the Helmholtz free energy. Similarly, we have analyzed all the
above properties for charged AdS BH with a global monopole.

We have deeply studied the influence of thermal corrections on
different important parameters of BH including mass of graviton $m$,
non-abelian hair $\nu$, non linear electrodynamics $\beta$. We
observed that critical horizons for the phase transitions shifted
due to the thermal corrections for both BHs and this range is
different from the usual range of phase transitions in literature
work \cite{46a}. This shifting is indicated by the specific heat and
Gibbs free energy for both the models. Our results proved that for
the large BHs, the impact of thermal corrections is negligible while
for small BHs, it has significant role. We have shown that for both
positive and negative cases, the behavior of specific heat and Gibbs
free energy has significant differences. Our study may be helpful to
investigate the deeply correlated condensed matter system, P-V
criticality, reentrant phase transitions, efficiency of heat engines
and Joule Thomson effect. Our results show that it is very important
to consider corrected thermodynamics to study the microscopic
interaction when black hole is small because quantum and thermal
fluctuations cannot be ignored.

\section*{Acknowledgment}
The author is thankful to HEC, Islamabad, Pakistan for its financial
support under the grant No: 9290/Balochistan/NRPU/R\&D/HEC/2017.

\end{document}